\documentclass[fleqn,usenatbib]{mnras}
\usepackage{newtxtext,newtxmath}
\usepackage[T1]{fontenc}
\DeclareRobustCommand{\VAN}[3]{#2}
\let\VANthebibliography\thebibliography
\def\thebibliography{\DeclareRobustCommand{\VAN}[3]{##3}\VANthebibliography}

\usepackage{graphicx}	
\usepackage{amsmath}

\usepackage[]{inputenc}
\usepackage{framed}
\usepackage{graphicx,epsfig,psfig,multicol}
\usepackage[dvipsnames]{xcolor}
\usepackage{amsmath}
\usepackage{mathtools}
\usepackage{color}
\usepackage{epsfig}
\usepackage{epstopdf}
\usepackage{textcomp}
\usepackage{xcolor,cancel}
\usepackage{comment}
\usepackage{subcaption}
\usepackage{array}
\usepackage{mathrsfs}

\usepackage[export]{adjustbox}
\usepackage{adjustbox}
\usepackage{placeins}

\usepackage{float}
\usepackage{morefloats}

\usepackage{lipsum}
\usepackage{mwe}
\usepackage{arydshln}
\usepackage{stmaryrd}

\usepackage{multirow}

\definecolor{webgreen}{rgb}{0,.5,0}
\definecolor{webbrown}{rgb}{.6,0,0}

\newcommand{\Msun}{\mbox{$ M_\odot $  }}

\title[Intermediate States]{Intermediate States in Chaotic Triple Evolution and Applications to Black Hole Merger Statistics}

\author[D. Meylakh, N.C. Stone \& N.W.C. Leigh]{
Dina Meylakh$^{1}$\thanks{E-mail: dina.meylakh@mail.huji.ac.il}
Nicholas C. Stone$^{1,2}$, 
Nathan W.C. Leigh$^{3}$
\\
$^{1}$Racah Institute of Physics, The Hebrew University, 91904, Jerusalem, Israel\\
$^{2}$Department of Astronomy, University of Wisconsin, Madison, WI 53706, USA\\
$^{3}$Departamento de Astronomia, Universidad de Concepcion
}

\date{Accepted XXX. Received YYY; in original form ZZZ}

\pubyear{\the\year{}}

\begin{document}
\label{firstpage}
\pagerange{\pageref{firstpage}--\pageref{lastpage}}
\maketitle

\begin{abstract}
Three-body interactions exhibit phases of strong chaotic evolution as well as hierarchical motion where one body separates from a binary and follows a hyperbolic or elliptic trajectory around it. The binaries produced during phases of hierarchical motion may lead to gravitational wave (GW) inspirals, but this depends on the outcomes of the chaotic states. 
In this paper we re-derive the elliptic outcome distribution using equilibrium statistical mechanics and explore it together with the hyperbolic distribution. 
When comparing to N-body simulations, we find that we can reduce the elliptic outcome model to one free parameter instead of the previously used two and that the predicted disintegration probabilities agree except for very low angular momentum triples. 
We then use both outcome distributions along with a star cluster model to design a Monte Carlo algorithm for repeated binary-single scatterings within dense star systems. 
We explore star cluster masses of \([10^5 - 10^7]M_{\odot}\), with the goal of quantifying observably eccentric merger (OEM) GWs, visible to instruments such as LIGO and Virgo.  Assuming an OEM detection sensitivity of \(f_{\rm min}=10 \text{Hz}, e_{\rm min} = 0.1\), we find the elliptic OEMs are about \(\sim (32 - 63)\%\) of the total elliptic mergers and that the total cluster mass greatly impacts the fraction of ejected binaries. 
The OEM to total merger fraction (OEM fraction) is found to be \((2.6 - 4.4)\%\). 
Considering the detection sensitivity that GW interferometers have today \((f_{\rm min} \simeq 34.4 \text{Hz})\) we obtain the OEM fraction in the  \((1.6 - 3.1)\%\) range.
\end{abstract}


\section{Introduction}
\label{sec:intro}

The search for an analytical solution to the problem of three gravitationally interacting bodies dates back to Newton. Although many forms of perturbation theory can be applied to hierarchical limits of the problem \citep{MurrayDermott99}, the motion of non-hierarchical triples evaded all early attempts at analytic modeling. It was not until two centuries later that Poincaré achieved a breakthrough by identifying the three-body problem's non-integrable essence, leading to the recognition of its unpredictability \citep{valtonen_three-body_2006}. This deterministic unpredictability can be found in many other dynamical systems and is now known as classical chaos. 

Non-hierarchical three-body motion exhibits phases of extremely chaotic behavior with rapid energy exchange between the bodies, 
along with phases of regular motion, where one can model their orbits
as two two-body systems: a temporary inner compact binary, and an outer system, composed of the tertiary orbiting the inner binary's center of mass \citep{heggie_binary_1975}.
We shall call the phases of extreme chaos \textit{scrambles} and the phases of predictable motion \textit{escapes} or \textit{excursions} depending on whether the tertiary is on an unbound orbit or a bound one, respectively. 

The study of the three-body problem is especially important in astrophysics, where it has implications for understanding binary-single scatterings within dense star systems. These interactions, where binaries encounter a third stellar object, play a crucial role in shaping stellar evolution and population dynamics within clusters\footnote{Binaries act as energy reservoirs. When binary-single encounters take place, their binding energy is transformed to kinetic energy, effectively heating up the cluster and expanding it if the singles are retained, or compressing it if the singles escape \citep{mapelli_maxwells_2018}.}. Such encounters often lead to the formation of unique systems, including binary stars with distinctive characteristics such as extreme eccentricities, mass transfer, or the formation of systems prohibited by isolated binary evolution.
Such binaries may lead to exotic sources such as 
cataclysmic variables (\citealt{hut_white_1983}, \citealt{pooley_dynamical_2006}) and X-ray binaries (\citealt{clark_x-ray_1975}, \citealt{hills_formation_1976}, \citealt{pooley_dynamical_2003}),
micro-tidal disruption events (\citealt{perets_micro-tidal_2016}, \citealt{kremer_hydrodynamics_2022}), 
the formation of blue stragglers in collisions (\citealt{leonard_stellar_1989},\citealt{leigh_where_2007},\citealt{leigh_analytic_2011}), 
and gravitational wave (GW) mergers \citep{portegies_zwart_black_2000}. 

At the time of writing, over 200 GW mergers have been observed by the LIGO-Virgo-KAGRA (LVK) collaboration, with the overwhelming majority originating from black hole binaries \citep{LIGO2025}. 
Numerous formation channels for these mergers have been proposed, including Roche-lobe overflow mass transfer (\citealt{olejak_impact_2021}, \citealt{gallegos-garcia_binary_2021}), common envelope evolution \citep{kruckow_common-envelope_2016}, hydrodynamic torques in the accretion discs of active galactic nuclei \citep{Stone+17, grishin_effect_2023} and the secular dynamics of hierarchical triples \citep{Antonini+17,Giri2026}.  Beyond these scenarios, the chaotic dynamics of non-hierarchical triples (which can be initiated by binary-single scatterings in various environments) are an additional and leading explanation for the LVK population of binary black holes (\citealt{rodriguez_post-newtonian_2018_formation}, \citealt{rodriguez_post-newtonian_2018_highly_eccentric}, \citealt{Samsing18}, \citealt{leigh_rate_2018} \citealt{fragione_origin_2020}). 

Given the large number of theoretical formation channels, a natural question emerges: how can we find conclusive imprints of any of these scenarios in limited GW data?  Most work to date has focused on mass, mass ratio, and spin distributions.  While it is now possible to divide observed binary black hole mergers into sub-populations identified by these variables \citep{Anarya+26}, theoretical predictions for population distributions are not yet precise enough to decisively associate empirical sub-populations with distinct formation channels.

An additional window into the problem of GW origins is eccentricity.  The ratio of binary mergers with detectable eccentricity to those of non-detectable eccentricity is estimated to be higher when the origin of the merger is of a three-body interaction \citep{Samsing18, romero-shaw_four_2022}. Out of the 90 mergers observed by the end of the O3 LVK run, 0-4 have detectable eccentricities (\citealt{abbott_search_2019}, \citealt{abbott_gwtc-3_2023}, \citealt{romero-shaw_four_2022}). This estimated ratio of eccentric to non-eccentric mergers, along with the growing number of observations, can, in turn, help in identifying the sources of the GW mergers being detected. 

To gain insight into the chaotic three-body motion and its role in GW signals, numerical simulations and explicit evolution of equations of motion (EOM) are usually employed (\citealt{AgekyanAnosova67}, \citealt{leigh_small-n_2012}, \citealt{leigh_illustrating_2018}, \citealt{samsing_assembly_2017}, \citealt{rodriguez_post-newtonian_2018_highly_eccentric}, \citealt{rodriguez_post-newtonian_2018_formation}, \citealt{samsing_mocca-survey_2018}, \citealt{di_carlo_merging_2019}, \citealt{dallamico_eccentric_2023}). Although these simulations can be done from first principles, they are often computationally expensive and are limited in the cluster variable ranges they can explore.
For most astrophysical applications, it is therefore desireable to have an analytic theory that can capture the outcomes of chaotic three-body scatterings.

For many applications, it suffices to characterize a micro-canonical ensemble of three-body systems rather than to solve for explicit equations of motion. 
This statistical approach invokes the ergodic hypothesis during a scramble, assuming that all possible microstates in phase space are equiprobable during and at the end of an individual scramble.  \citealt{monaghan_statistical_1976} initially introduced this formalism for systems of low angular momentum, and for technical reasons did not require its conservation. This approach was subsequently developed to deal with conservation of scalar angular momentum (\citealt{monaghan_statistical_1976-1}, \citealt{nash_statistical_1978}, \citealt{valtonen_three-body_2006} chapter 7). It was not until \citealt{StoneLeigh19} simplified the problem using Delaunay elements, that a general analytical expression of the distribution of binary properties for escapes was achieved (here general means that all angular momentum components are conserved). A similar treatment of excursions by \citealt{GinatPeretz2021} closely followed, although instead of an analogous orbital binary properties outcome distribution, a distribution as a function of angular momenta components was presented. A different approach describing phase-volume flux instead of phase volumes was also initiated and extensively studied by \citealt{kol_flux-based_2021}, \citealt{kol_natural_2021}, \citealt{manwadkar_testing_2021}, and \citealt{dandekar_regularized_2022}.  

In this paper, we derive the 3d outcome distribution of binary orbital elements for excursions (\S \ref{sec:derivation}), completely analogously to the escape distribution in \citealt{StoneLeigh19}. We then discuss the relevant phase space for integration (\S \ref{sec:excursions}), and the distribution features along with its marginalization (\S \ref{sec:features}). Further analytical integration, reducing the distribution dimensionality to 2, is derived in (App. \ref{app:CB}). Once we have distributions for escapes and excursions, we move to discuss disintegration probabilities and compare probabilities of scramble numbers to numerical 3-body simulations (\S \ref{sec:lifetimes}). We then set out to find the ratio of eccentric mergers to non-eccentric mergers as outcomes of repeated three-body interactions in globular clusters by sampling the relevant distributions. First, we describe a two-component cluster framework that we use to model the system's environment and initial properties (\S \ref{sec:cluster_dist}). Next, we describe the binary properties for physical exotic phenomena to be considered to occur (\S \ref{sec:binary_props_for_diss}) and proceed to describe repeated binary-single scattering evolution (\S \ref{sec:repeated_scatterings}). Finally, in \S \ref{sec:results} we discuss the results of the sampling simulation and derive an approximate analytical expression for a cluster mass above which no ejected binaries are possible. We outline our conclusions in \S \ref{sec:conclusions} and comment on the code availability at the end.

\section{Intermediate States}
\label{sec:IMS}

In this section, we compute phase volumes associated with ``intermediate states'' of the chaotic three-body problem, during which an inner binary evolves subject to perturbations from a bound but temporarily distant tertiary.  We use these phase volumes, which were first presented in \citet{GinatPeretz2021}, to estimate distributions of outcomes (e.g. binary orbital parameters) and to further investigate properties of intermediate states. 

\subsection{Derivation of the Elliptic Outcome Distribution}
\label{sec:derivation}
Let 
\(m_{\rm a}\), \(m_{\rm b}\) be the masses of the two bodies composing the binary,
\(m_{\rm B} = m_{\rm a} + m_{\rm b}\) the binary mass, \(m_{\rm s}\) the mass of the tertiary,
\(M\) the total mass, \(m\) the reduced mass of the binary-single system and \(\mathcal{M}\) the reduced mass of the binary, as detailed in Eq. \ref{eq:masses}.

\begin{equation}
M=m_{\rm s}+m_{\rm B}\mspace{50mu}m=\frac{m_{\rm B}m_{\rm s}}{M}\mspace{50mu}\mathcal{M}=\frac{m_{\rm a}m_{\rm b}}{m_{\rm B}}
\label{eq:masses}
\end{equation}
We define \(r_{\rm B}, p_{\rm B}\) as the relative position and momentum of the binary, and 
\(r_{\rm s}, p_{\rm s}\) the position and momentum of the tertiary relative to the binary's center of mass.  We further define
\(a\) to be the semimajor axis,
\(e\) the eccentricity,
\(I\) the inclination,
\(\lambda\) the mean anomaly,
\(\omega\) the longitude of pericenter,
\(\Omega\) the longitude of ascending node, and
\(\epsilon\) the eccentric anomaly.

In order to integrate over the available phase space volume, we make use of the canonical Delaunay variables defined as:
\begin{align}
    \Lambda = \sqrt{G\Tilde{m}a} \mspace{150mu}
    \lambda = \lambda \notag\\
    \Gamma = \sqrt{G\Tilde{m}a\left(1-e^{2}\right)} \mspace{150mu}
    \gamma = \omega \notag\\
    H = \sqrt{G\Tilde{m}a\left(1-e^{2}\right)}\cos{I} \mspace{150mu}
    \eta = \Omega
    \label{eq:delaunay}
\end{align}
We add a subscript \({\rm B}\) and use \(\Tilde{m}=m_{\rm B}\) to indicate the binary's orbital elements, while adding a subscript of \({\rm s}\) and using \(\Tilde{m}=M\) for the binary-single orbital elements.

We define \(R\) to be the radial position of the boundary of the chaotic region relative to the binary's center of mass; following previous work \citep{Valtonen05, StoneLeigh19, GinatPeretz2021}, we assume that when all three bodies are within a radial separation of $R$, the orbits are strongly chaotic, but when the tertiary is at a greater distance from the binary center of mass, the motion can decouple into the quasi-regular evolution of two perturbed two-body problems.  This assumption (which we shall return to later) is clearly an idealization, although it agrees with numerical simulations well for physically motivated choices of $R$; see however \citet{kol_flux-based_2021} for an alternative perspective.

We calculate the value of the tertiary's mean anomaly \(\lambda_{\rm s}\) as a function of its position \(r_{\rm s}\) using the eccentric anomaly $\epsilon_{\rm s}$ and the coordinate relations \(r_{\rm s} = a_{\rm s}\left(1-e_{\rm s}\cos{\epsilon_{\rm s}}\right)\) and \(\lambda_{\rm s} = \epsilon_{\rm s} -e_{\rm s}\sin{\epsilon_{\rm s}}\).
This results in the following expression:
\begin{equation}
\lambda_{\rm s}\left(r_{\rm s}\right) = \cos^{-1} \left( \frac{1-\frac{r_{\rm s}}{a_{\rm s}}}{e_{\rm s}} \right)-\sqrt{e_{\rm s}^2 -1 + \frac{2r_{\rm s}}{a_{\rm s}} - \frac{r_{\rm s}^2}{a_{\rm s}^2}}, \label{eq:chaoticR}
\end{equation}
recovering Eq. 24 in \citep{GinatPeretz2021}.
Substituting \(r_{\rm s} = R\) and using Delaunay variables in the latter expression, results in the value of the tertiary's mean anomaly at the boundary of the chaotic region:
\begin{equation}
\lambda_{\rm s}\left(r_{\rm s}=R\right) = \cos^{-1} \left( \frac{1-\frac{GMR}{\Lambda_{\rm s}^{2}}}{\sqrt{1-\frac{\Gamma_{\rm s}^2}{\Lambda_{\rm s}^2}}} \right)-\sqrt{\frac{2GMR}{\Lambda_{\rm s}^{2}} -\frac{\Gamma_{\rm s}^2}{\Lambda_{\rm s}^2} -\frac{G^2M^2R^2}{\Lambda_{\rm s}^{4}}}
\end{equation}
We are now able to integrate over the available phase space volume:
\begin{equation}
\sigma = \int\dots\int\delta\left(E_{\rm B}+E_{\rm s}-E_{\rm 0}\right)\delta\left(\vec{L_{\rm B}}+\vec{L_{\rm s}}-\vec{L_{\rm 0}}\right){\rm d}\vec{r_{\rm B}}{\rm d}\vec{p_{\rm B}}d\vec{r_{\rm s}}{\rm d}\vec{p_{\rm s}}
\label{eq:integral}
\end{equation}
We transform the Cartesian coordinates in Eq. {\ref{eq:integral}} to Delaunay variables, denoting
\(\vec{D}_{\rm B}\) as the binary Delaunay variables and
\(\vec{D}_{\rm s}\) as the tertiary Delaunay variables:
\begin{align}
    &\sigma = \int\dots\int\delta\left(-\frac{G^2m_{\rm a}m_{\rm b}m_{\rm B}}{2\Lambda_{\rm B}^2}-\frac{G^2m_{\rm B}m_{\rm s}M}{2\Lambda_{\rm s}^2}-E_{\rm 0}\right)\times \nonumber \\
    &\delta\left(\mathcal{M}H_{\rm B}+mH_{\rm s}-L_{\rm 0}\right)\delta\left(\mathcal{M}\Gamma_{\rm B}\sin{\eta_{\rm B}}\sin{I_{\rm B}}+m\Gamma_{\rm s}\sin{\eta_{\rm s}}\sin{I_{\rm s}}\right) \nonumber\\
    &\times\delta\left(\mathcal{M}\Gamma_{\rm B}\cos{\eta_{\rm B}}\sin{I_{\rm B}}+m\Gamma_{\rm s}\cos{\eta_{\rm s}}\sin{I_{\rm s}}\right)d\vec{D}_{\rm B}d\vec{D}_{\rm s}.
\end{align}
We integrate \(\lambda_{\rm B}, \gamma_{\rm B}, \gamma_{\rm s}\) across \(\{0, 2\pi\}\), and \(\lambda_{\rm s}\) from 0 to \(\lambda_s\left(r_{\rm s}=R\right)\). We then integrate over the Dirac delta functions to eliminate the tertiary's orbital elements. Finally, we transform the Delaunay variables back to Keplerian orbital elements and find the simple 3d phase volume for intermediate states in the chaotic three-body problem, subject to energy and angular momentum conservation:
\begin{align}
&\sigma = \frac{\pi^4G^4M^{\frac{5}{2}}\left(m_{\rm a}m_{\rm b}\right)^{\frac{3}{2}}}{m_{\rm s}^{\frac{3}{2}}}\int\int\int\frac{e_{\rm B}{\rm d}E_{\rm B}{\rm d}e_{\rm B}{\rm d}C_{\rm B}}{L_{\rm s}\left(E_{\rm B}-E_{\rm 0}\right)^{\frac{3}{2}}\left(-E_{\rm B}\right)^{\frac{5}{2}}} \nonumber \\
&\times\left(\cos^{-1}\left(\frac{1-\frac{2R}{Gm_{\rm B}m_{\rm s}}\left(E_{\rm B}-E_{\rm 0}\right)}{\sqrt{1-\frac{2M\left(E_{\rm B}-E_{\rm 0}\right)L_{\rm s}^2}{G^2m_{\rm B}^3m_{\rm s}^3}}}\right)\right. \nonumber\\
&\left.-\sqrt{\frac{2M\left(E_{\rm B}-E_{\rm 0}\right)}{G^2m_{\rm s}^3m_{\rm B}^3}\left(2RGMm^2-2m\left(E_{\rm B}-E_{\rm 0}\right)R^2-L_{\rm s}^2\right)} \right).
\label{eq:3ddist}
\end{align}
Here \(C_{\rm B} = \cos{I_{\rm B}}\) and the tertiary angular momentum
\begin{align}
L_{\rm s}^2=L_{\rm B}^2\left(1-C_{\rm B}^2\right)+\left(L_{\rm B}C_{\rm B}-L_{\rm 0}\right)^2 \label{eq:Ls}
\end{align}
is used for short.  

Assuming an ergodic exploration of the three-body problem's phase volume \citep[i.e. following the original proposal of][]{monaghan_statistical_1976}, the integrand of Eq. \ref{eq:3ddist} represents the elliptic (or intermediate state) {\it outcome distribution}.  In other words, this integrand is the differential distribution of binary parameters in a hypothetical microcanonical ensemble of many three-body systems, all of which share the same $E_0$, $\vec{L}_0$, and mass triplet $\{ m_{\rm a}, m_{\rm b}, m_{\rm s} \}$, as evaluated during individual intermediate states.  This distribution was first derived in \citet{GinatPeretz2021} and parallels the analogous hyperbolic (or final state) outcome distribution (i.e. binary properties following ejection of the tertiary onto a hyperbolic rather than elliptical orbit) from \citet{StoneLeigh19}.

While we did not find a way to perform the full triple integral for the phase volume $\sigma$ analytically, it is possible to analytically integrate over the cosine-inclination $C_{\rm B}$; the resulting bivariate marginal distribution (in $E_{\rm B}$ and $L_{\rm B}$) is presented in Appendix \ref{app:CB}.

\subsection{Constraints on Excursions}
\label{sec:excursions}
Given various initial energies, masses and angular momenta, integrating Eq. \ref{eq:3ddist} within the naive intermediate state boundaries 
\begin{align}
E_{\rm B} \in [E_{\rm 0}, 0],  C_{\rm B} \in [-1, 1],  e_{\rm B} \in [0, 1]
\label{eq:integrationBoundsNaive}
\end{align}
 often yields a complex-valued phase volume. Non-real components of the phase volume result from regions in the \(\left(E_{\rm B}, C_{\rm B}, e_{\rm B}\right)\) phase space that cannot simultaneously (i) produce a tertiary in the chaotic region and (ii) conserve all integrals of motion.
For example, the basic definition of a bound excursion (\(0 < e_{\rm s} < 1\) ,  \(E_{\rm s} < 0\)) provides a constraint between the tertiary's angular momentum and its energy: 
\begin{equation}
L_{\rm s}^2 < - \frac{G^2M^2m^3}{2E_{\rm s}}.
\label{eq:LsEs}
\end{equation}
Adding in the demand for angular momentum conservation (Eq. \ref{eq:Ls}) and using energy conservation to rewrite \(E_{\rm s}\) results in new boundaries for \(L_{\rm B}\), which must lie between the two roots
\begin{equation}
L_{\rm B}^{\pm} = L_{\rm 0}C_{\rm B} \pm \sqrt{L_{\rm 0}^2 \left(C_{\rm B}^2 -1\right) - \frac{G^2M^2m^3}{2\left(E_{\rm 0} - E_{\rm B}\right)}}.
\label{eq:LBbounds}
\end{equation}
This reasoning motivates us to find further physical constraints that may limit the relevant phase space volume.
As mentioned earlier, we posit that in order for the system to be chaotic, the tertiary's pericenter \(q_{\rm s}\) should be inside the chaotic region:
\begin{equation}
q_{\rm s} = a_{\rm s}\left(1-e_{\rm s}\right) \le R.
\label{eq:qslessR}
\end{equation}
Rewriting in terms of angular momentum and energy results in
\begin{equation}
L_{\rm s}^2 \le 2RGMm^2 + 2E_{\rm s}R^2m.
\label{eq:Lscrit}
\end{equation}
Because \(0 < 2mR^2\) the latter equation can be rewritten as
\begin{equation}
 E_{\rm crit} \equiv - \frac{GMm}{R} + \frac{L_{\rm s}^2}{2mR^2} \le E_{\rm s}.
 \label{eq:Ecrit}
\end{equation}
This can be interpreted as demanding that the tertiary's energy is larger than \(E_{\rm crit}\), its energy when \(r_{\rm s}=R\).
Using conservation laws we can also write  Eq. \ref{eq:Ecrit} in terms of the binary properties:
\begin{equation}
L_{\rm 0}^2 - 2L_{\rm B}L_{\rm 0}C_{\rm B} + L_{\rm B}^2 \le 2RGMm^2 + 2\left(E_{\rm 0} - E_{\rm B}\right)R^2m.
\label{eq:StartForBounds}
\end{equation}
We note Eqs. \ref{eq:Lscrit} and \ref{eq:StartForBounds} are mathematically equivalent to requiring that (i) the radical in Eq. \ref{eq:3ddist} is non-negative, and equivalently that (ii) the argument of the \(\cos^{-1}\) in Eq. \ref{eq:3ddist} lies between -1 and 1.  In other words, requiring a real-valued phase volume can be physically interpreted as demanding that outcomes in intermediate states (a) conserve $E_0$, (b) conserve $\vec{L}_0$, and (c) have tertiary pericenters $q_{\rm s} \le R$.

For practical evaluation of outcome distributions, the addition of a Heaviside function
encapsulating Eq. \ref{eq:StartForBounds} removes the imaginary parts from the results. For computational efficiency, we do not use a Heaviside function but instead employ exact analytic phase space bounds as described in Appendix \ref{app:phasespacebounds}.

Another limit on the phase space volume -- irrelevant in the hyperbolic case -- is the tertiary's semi-major axis. In order for the binary to be clearly separable from the tertiary and for a quasi-regular excursion to form, the semi-major axis of the tertiary must be larger than the semi-major axis of the binary by at least some amount. We therefore demand that 
\begin{equation}
    \beta a_{\rm B} \leq a_{\rm s}
    \label{eq:betta_def}
\end{equation}
for some dimensionless parameter \(\beta > 1 \).
This is different than what was previously done in \cite{GinatPeretz2021}, where the hierarchy was imposed on the energies rather than the semi-major axis. We prefer to work with this critrea instead to allow identification of the closest two objects as the binary, regardless of their masses.
This in turn can be rewritten as a constraint on the binary energy:
\begin{equation}
E_{\rm B} \leq \frac{\beta m_{\rm a}m_{\rm b}E_{\rm 0}}{m_{\rm s}m_{\rm B}+\beta m_{\rm a}m_{\rm b}} \equiv E_{\rm cut}
\label{eq:EBcut_def}
\end{equation}

Following \citet{StoneLeigh19} we will consider two different definitions for the chaotic region \(R\). The \textit{simple escape} (SE) criterion where 
\begin{equation}
R = \alpha a_{\rm B} 
\label{eq:simpleesc}
\end{equation}
and the \textit{apocentric escape} (AE) criterion where 
\begin{equation}
R = \alpha a_{\rm B}\left(1 + e_{\rm B}\right)
\end{equation}
invoke a dimensionless parameter \(\alpha\) for both cases. 
Later in the paper, comparison with numerical scattering experiments (\S \ref{sec:lifetimes}) motivates us to use the value \(\alpha = 2.5\) for the SE criteria and \(\alpha = 3.5\) for AE.  Interestingly, our later numerical experiments find that \(\beta = \alpha \) for both cases, perhaps suggesting that both fudge factors may have a common physical origin in e.g. the triple stability boundary \citep{MardlingAarseth01, LalandeTrani22}.

\subsection{Features of the Elliptic Outcome Distribution}
\label{sec:features}
\begin{figure}
\includegraphics[width=85mm]{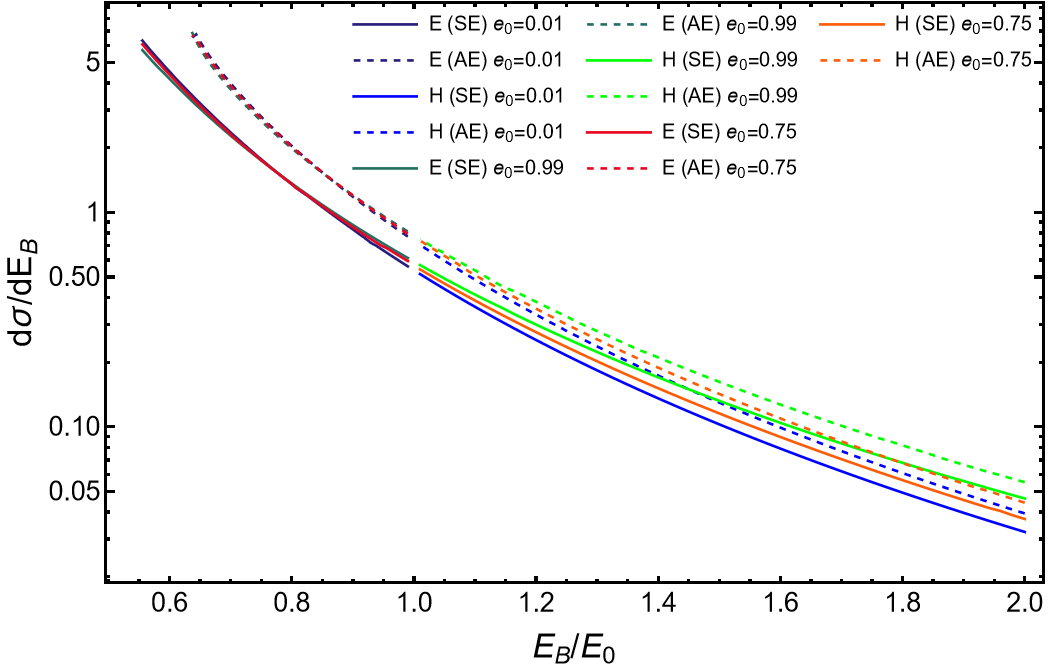}
\caption{The marginal distribution of binary energy, \({\rm d}\sigma/{\rm d}E_{\rm B}\) for both hyperbolic ($E_{\rm B}/E_0 > 1$) and elliptic ($E_{\rm B}/E_0 < 1$) tertiary orbits, plotted against the dimensionless energy \(E_{\rm B}/E_{\rm 0}\). The dotted lines correspond to the apocentric escape (AE) chaotic region, and filled lines to the simple escape (SE) chaotic region. The colors correspond to the systems different angular momenta $L_0$ (here represented with $e_0$), including a high (blue), a medium (red and orange) and a low (green) $L_0$ case. The light colors are used for the hyperbolic outcomes (light blue, light green and orange) while the darker colors (blue, red, green) are used for elliptic outcomes. The elliptical binary energy distribution is a smooth continuation of the hyperbolic case, with a shared \(E_B = E_0\) limit predicted in \ref{eq:E0limit}, and the system appears to be insensitive to changes in initial angular momenta for both escape criteria.  The slight difference in normalization between the AE and SE curves reflects the small difference in the minimum $E_{\rm B}$ of an intermediate-state binary.}
\label{fig:Energy1Ddist}
\end{figure}

As a consistency check for Eq. \ref{eq:3ddist}, we take the \(E_B \rightarrow E_0\) limit of the integrand, applying L'H\^{o}pital's rule. In this limit, the energy of the tertiary approaches 0 and therefore its orbit becomes nearly parabolic:
\begin{align}
\lim_{E_B \rightarrow E_0}&\frac{ {\rm d}\sigma}{{\rm d}E_{\rm B}{\rm d}e_{\rm B}{\rm d}C_{\rm B}} =  \frac{8\pi^4 M^{7/2}}{3G\left(m_{\rm a} m_{\rm b} \right)^{3/2} m_{\rm s}^{11/2}m_{\rm B}^3}  \notag\\
&\times \frac{L_B}{\left(-E_0\right)^{3/2}L_{\rm s}} \frac{2G^2m^3M^2R^2+GMmL_s^2R-L_s^4/m}{\sqrt{4Gm_sm_BR-2L_s^2/m}}
\label{eq:E0limit}
\end{align}
This result is identical to the equivalent parabolic limit of the hyperbolic outcome distribution. (\citealt{StoneLeigh19} Eq. 21).
This can be seen graphically in the equal-mass, marginalized 1d energy distribution, ${\rm d}\sigma/{\rm d}E_{\rm B}$, as shown in Fig. \ref{fig:Energy1Ddist}. For this figure, we integrated Eq. \ref{eq:3ddist} over \(C_{\rm B} \in [-1, 1], e_B \in [0, 1] \) for both the elliptic and the hyperbolic cases, considering multiple cases with different values for $L_{\rm 0}$ and definitions for the strong interaction region $R$. The variation over the angular momentum \(L_{\rm 0}\) is encoded by the dimensionless ``effective eccentricity'' \(e_{\rm 0}\) parameter, where \(L_{\rm 0}=\mathcal{M}\sqrt{-G^{2}m_{\rm B}m_{\rm a}m_{\rm b}\left(1-e_{\rm 0}^{2}\right)/2E_{\rm 0}}\).
We see that in all cases there is a smooth continuation through $E_{\rm B} = E_0$, joining the elliptic and hyperbolic outcome distributions.  This reflects the absence of a qualitative physical difference between a scramble that produces a marginally hyperbolic ejection from one that produces a marginally elliptical ejection; mathematically, this continuity was already visible in Eq. \ref{eq:E0limit}.  We also find that the differential outcome distributions ${\rm d}\sigma/{\rm d}E_B$ depend very weakly on $L_0$ and the definition of $R$ (although for highly hyperbolic outcomes with $E_{\rm B} / E_0 \gg 1$, $L_0$ affects the location of a cutoff in the distribution; \citealt{StoneLeigh19}).  

\begin{figure}
\includegraphics[width=85mm]{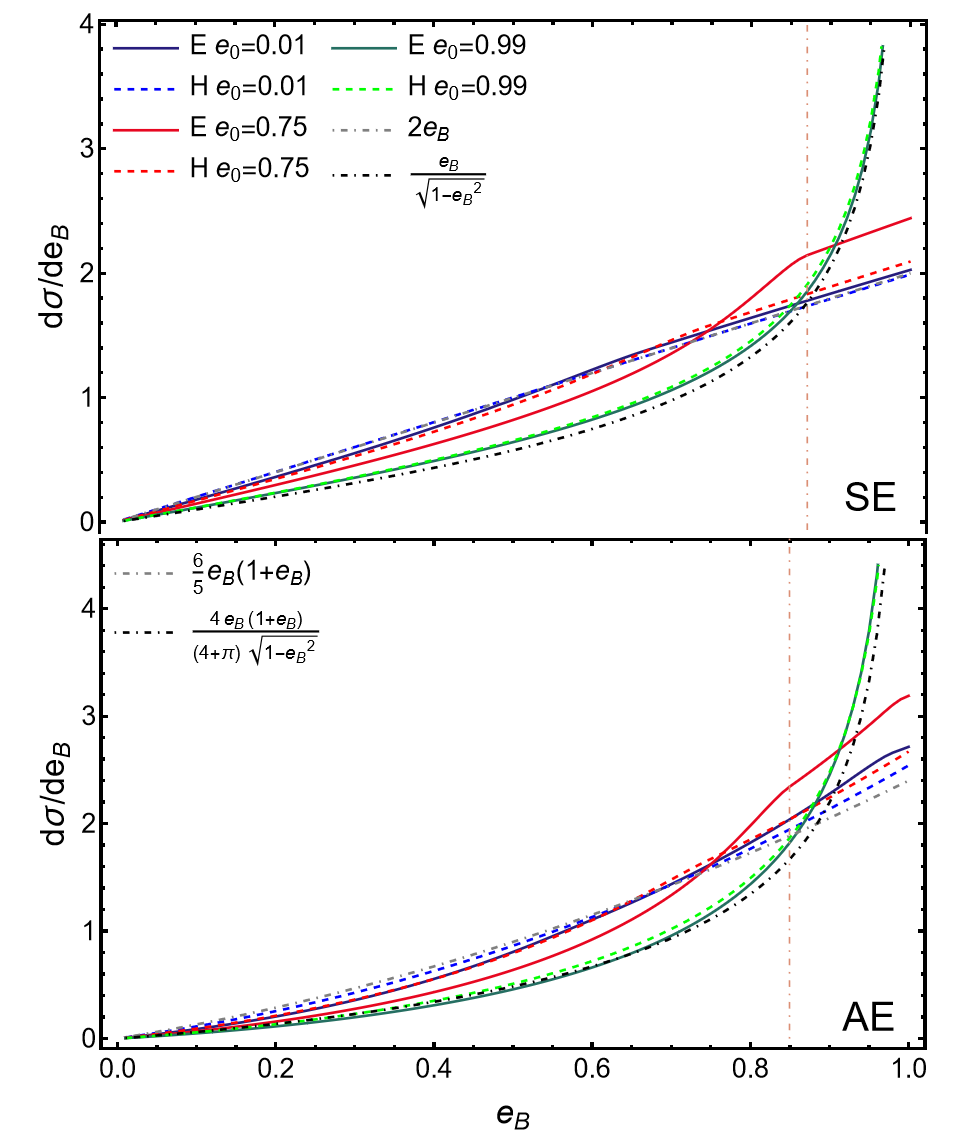}
\caption{The normalized marginal distribution of binary eccentricity, \({\rm d}\sigma/{\rm d}e_B\) for both hyperbolic and elliptic tertiary orbits, plotted against eccentricity \(e_B\). The dotted curves in both panels correspond to hyperbolic distributions while the filled curves correspond to elliptic distributions. The bottom panel corresponds to the apocentric escape (AE) chaotic region, and the top panel to the simple escape (SE) chaotic region. Limiting super-thermal distributions for both chaotic region criteria are in dot-dashed black, and thermal distributions are in grey. The other colors correspond to a range of different angular momenta, from high (blue, light blue), to medium (red and orange) and finally low (green, light green). The distributions for high angular momenta are roughly thermal while for low angular momenta they are super-thermal, for both elliptical and hyperbolic cases. For mid-range total angular momenta a second order discontinuity can be seen clearly for the elliptic distribution in both panels, predicted in Eq. \ref{eq:Spike}, and plotted as vertical lines with \(E_{\rm B}\) taken as in Eq. \ref{eq:EBcut_def}.}
\label{fig:eBdistDividedtoAESE}
\end{figure}

Likewise, we marginalize over \(E_B \in [E_0, E_{\rm cut}]\), and \(C_B \in [-1, 1] \) for the 1d eccentricity distribution \({{\rm d}\sigma}/{{\rm d}e_{\rm B}} \) and produce Fig. \ref{fig:eBdistDividedtoAESE}. We see a higher density of states for higher binary eccentricity values \(e_{\rm B}\), with properties quite similar to the hyperbolic case \citep{StoneLeigh19}. We summarize the main results of these calculations as follows:
\begin{itemize}
    \item We find a quasi-thermal distribution for higher $L_0$ values, which is precisely thermal (${\rm d}\sigma / {\rm d} e_{\rm B} = 2e_{\rm B}$ with $\left\langle e_{\rm B} \right\rangle = 2/3$) for the SE criterion, and mildly super-thermal (${\rm d}\sigma / {\rm d} e_{\rm B} =(6/5)(1+e_{\rm B})e_{\rm B}$  with $\left\langle e_{\rm B} \right\rangle = 0.7$) for the AE criterion.
    \item We find highly super-thermal distributions as $L_0 \to 0$, with ${\rm d}\sigma / {\rm d} e_{\rm B} = e_{\rm B}/ \sqrt{1-e_{\rm B}^2}$   and $\left\langle e_{\rm B} \right\rangle = 0.8$ for the SE criterion and ${\rm d}\sigma / {\rm d} e_{\rm B} = 4 (4+\pi)^{-1} e_{\rm B}(1+e_{\rm B})/ \sqrt{1-e_{\rm B}^2}$   with $\left\langle e_{\rm B} \right\rangle \simeq 0.83$ for the AE criterion.
\end{itemize} 
Both these results parallel equivalent hyperbolic outcome distributions \citep{StoneLeigh19}.  We also see a discontinuity in the first derivative, which is most apparent for intermediate angular momentum values.

This discontinuity reveals another physical feature of the outcome distribution, which is the presence of \(L_s\) in the denominator. This implies higher values for the probability density function when \(L_s \rightarrow 0\). From angular momentum conservation, we know \(L_s = 0\) could happen only when \(C_B = 1\). If we rewrite \(L_B\) in terms of binary eccentricity and energy we can predict a 2d divergence in the \( \{ E_B, e_B, C_B = 1 \} \) phase space. Specifically, the divergence should occur at:
\begin{align}
e_B = \sqrt{1 + \frac{2E_B L_0^2}{G^2 m_a m_b m_B \mathcal{M}^2}}
\label{eq:Spike},
\end{align}
which is in reasonable agreement with results in Fig. \ref{fig:eBdistDividedtoAESE}.
This divergence originates from an elimination of a mutual constraint between the binary energy and angular momentum (recall Eq. \ref{eq:LsEs},  \ref{eq:LBbounds}), causing their values to span independent and broader regions of phase space. We use this discontinuity to analytically approximate \(d\sigma/de_{\rm B}\) in App. \ref{app:eBapproximation}.

\begin{figure}
\centering
\includegraphics[width=85mm]{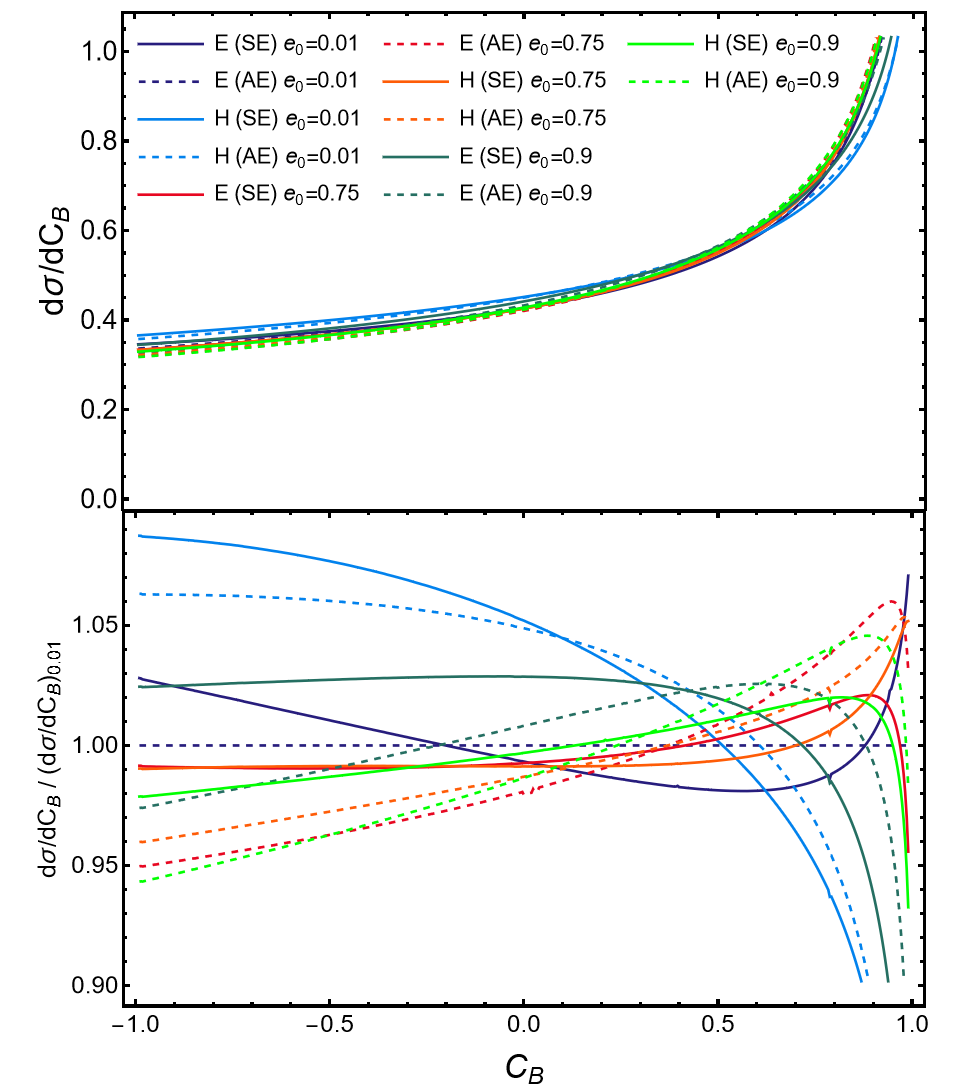}
\caption{The normalized marginal distribution of binary orientation, \({\rm d}\sigma/{\rm d}C_B\) for both hyperbolic and elliptic tertiary orbits, plotted against orientation \(C_B\). The dotted and filled lines and their colors are the same as Fig. \ref{fig:Energy1Ddist}. In the bottom panel, all distributions were normalized in relation to the elliptic AE low angular momentum case to highlight their point-wise differences, which do not deviate from 10\% for most of the $C_{\rm B}$ domain.}
\label{fig:1ddistCBdividedbyAEe001}
\end{figure}

We further marginalize over \(E_{\rm B} \in [E_0, E_{\rm cut}]\), and \(e_{\rm B} \in [0, 1] \) for the 1d cos-inclination distribution \({{\rm d}\sigma}/{{\rm d}C_{\rm B}} \) and produce Fig. \ref{fig:1ddistCBdividedbyAEe001}. We see a higher density of states as \(C_{\rm B}\) gets closer to 1. Compared to ${\rm d}\sigma/{\rm d}E_{\rm B}$ and ${\rm d}\sigma/{\rm d}e_{\rm B}$, ${\rm d}\sigma/{\rm d}C_{\rm B}$ exhibits far less dependence on both \(L_{\rm 0}\) and on the chaotic region criteria.  No equal-mass results deviate from the AE elliptic, high angular momentum outcome distribution by more than \(10\%\) for most of the region.


\section{Non-Hierarchical Triple Lifetime Distributions}
\label{sec:lifetimes}

Having explored the {\it continuous} outcome variables of three-body intermediate states, in this section, we move on to examine the {\it discrete} outcome variables that shape their properties and, at a more fundamental level, determine relative probabilities of bound excursions and unbound escapes.  More specifically, the discrete outcome variables we examine here are a set of branching ratios that will be computed from the same ergodic assumptions motivating the previous section of this paper.  We will also combine these branching ratios in a novel way to predict the lifetime distributions of metastable, non-hierarchical triples.

We define \(\sigma_{\rm ex} \) as the phase space volume corresponding to an elliptic excursion, meaning the phase space bounded maximally \footnote{As described in Eq. \ref{eq:StartForBounds}, the actual phase space volume that can be occupied by intermediate states may be substantially smaller than these maximal limits; the same reasoning applies also to hyperbolic outcome states \citep{StoneLeigh19}.}  by \(e_B\in [0, 1]\), \(C_B \in [-1, 1]\) and \(E_B \in [E_0, E_{\rm cut}] \). Likewise, the phase space volume corresponding to an escape \(e_B \in [0, 1]\), \(C_B \in [-1, 1]\), \(E_B \in [-\infty, E_0]\), will be denoted \(\sigma_{\rm es}\). 
For a given 3-body mass triplet \(m_1, m_2, m_3\), we will add the superscript \(1,2,3\) to \(\sigma_{\rm ex}, \sigma_{\rm es}\) to indicate that \(m_{\rm 1}, m_{\rm 2}\) are the binary masses and \(m_{\rm 3}\) is the tertiary mass. Given an intermediate state excursion, the ergodic hypothesis would yield a probability that the object of mass \(m_{\rm 3}\) is the ejected tertiary is:

\begin{align}
 P (m_3\;ex| ex) = \frac{\sigma_{\rm ex}^{\rm 1,2,3}}{\sigma_{\rm ex}^{\rm 1,2,3} + \sigma_{\rm ex}^{\rm 2,3,1} + \sigma_{\rm ex}^{\rm 3,1,2}}
 \label{eq:Pm3ex}
\end{align}
Similarly, the ergodic probability of an \(m_3\) escape given an escape from the system (where any mass can escape) will be
\begin{align}
 P (m_3\;es| es) = \frac{\sigma_{\rm es}^{\rm 1,2,3}}{\sigma_{\rm es}^{\rm 1,2,3} + \sigma_{\rm es}^{\rm 2,3,1} + \sigma_{\rm es}^{\rm 3,1,2}}
 \label{eq:Pm3es}
\end{align}
The same probabilities can be computed for excursions and escapes of the other two particles by permutation of indices.

\begin{figure}
\includegraphics[width=80mm]{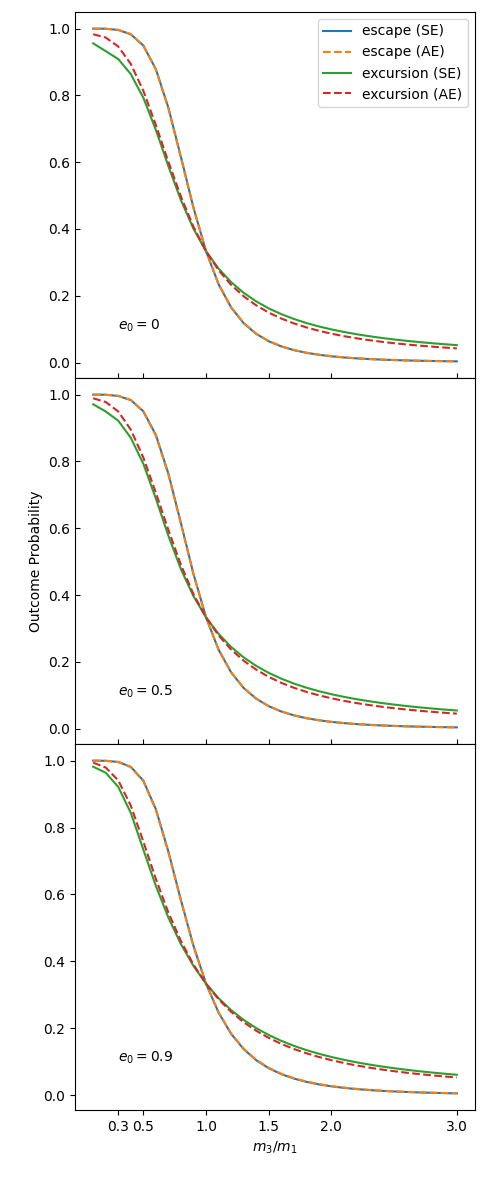}
\caption{Probability of \(m_3\) being selected as the tertiary to undergo an escape or excursion, given any escape or excursion in the system. We consider the \(m_1 = m_2 \) case and plot probabilities against the \(m_3/m_1\) mass ratio. Probabilities of \(m_s\) escape given an escape are shown in blue for the SE criteria, and in orange for AE. Probabilities of \(m_s\) excursion given an excursion are shown in green for the SE criteria, and in red for AE. The three panels correspond to different total angular momenta of the 3-body system, starting from the highest at the top, and lowest at the bottom. For \(m_{\rm 3}/m_{\rm 1} = 1\) we recover the equal mass case, and all event probabilities \(=1/3\). The probability for an \(m_3 < m_1\) to escape given an escape are larger than for it to have an excursion given an excursion, and the opposite for \(m_3 > m_1 \).}
\label{fig:EscapeExcursionProb}
\end{figure}

\begin{figure}
\includegraphics[width=80mm]{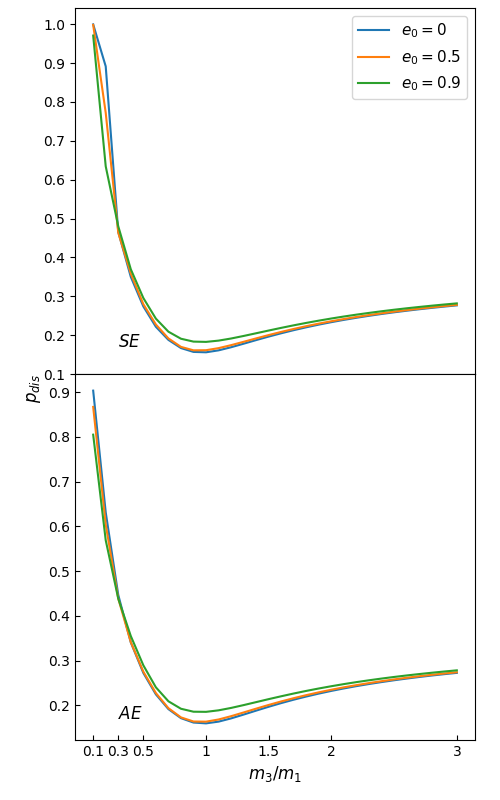}
\caption{Probability of an escape given any exit (either escape or excursion) from the chaotic region, taken with \(m_1 = m_2\) and plotted against the \(m_3/m_1\) mass ratio. Colors denote the total angular momentum \(L_{\rm 0}\), varied from lowest (green) through intermediate (orange) to highest (blue). The top panel corresponds to the simple escape (SE) and the bottom panel to the apocentric escape (AE) definition of the chaotic region $R$. We see that for all starting angular momenta, the most ``stable'' (i.e. least prone to disintegration) mass configuration is around the equal mass point. Lower angular momentum generally increases the disintegration probability for all mass and chaotic region configurations, though the effect is modest.}
\label{fig:DisintegrationProb}
\end{figure}

\begin{figure}
\includegraphics[width=80mm]{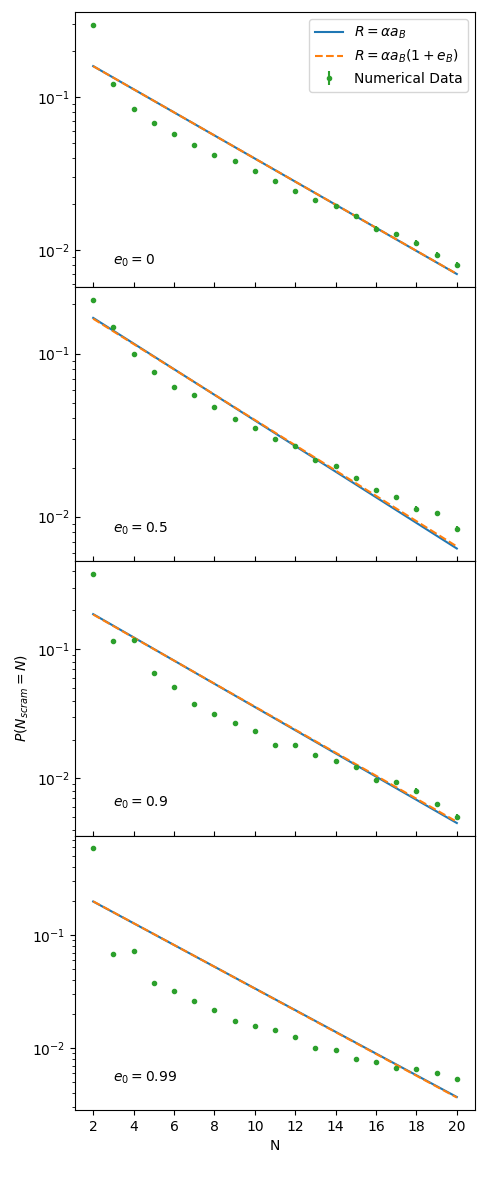}
\caption{Probability of a system to exhibit a certain number of scrambles, \(N_{\rm scram}\), plotted against \(N_{\rm scram}\). Calculations with the simple escape (SE) criteria are shown in blue, and for the apocentric escape (AE) in orange. Data gathered from the binary-single numerical scattering experiments of \citet{StoneLeigh19} is shown in green. The four panels correspond to different total angular momenta of the 3-body system, starting from the highest at the top, and lowest at the bottom. The sum of probabilities was normalized to 1, starting at \(N_{\rm scram} = 2\) for all data sets. 
We see quite good agreement for $N_{\rm scram}>2$ for high $L_0$ systems.  The agreement is worse for low $L_0$ and low $N$ values, though it generally improves with increasing $N$.  In the most extreme case we examine ($e_0=0.99$), the disagreement between our analytic predictions and numerical simulations rises to a factor of a few.}
\label{fig:RandAEvsNumerical}
\end{figure}

In Fig. \ref{fig:EscapeExcursionProb} we plot equations \ref{eq:Pm3ex} and \ref{eq:Pm3es} with an equal mass pair \(m_{\rm 1} = m_{\rm 2}\) while varying the third mass \(m_{\rm 3}\) and the total angular momentum $L_0$ of the system. One immediate conclusion from this figure is that these conditional branching ratios are almost completely insensitive to our choice of approximation for the chaotic region.  When $m_3 < m_1$, it is favored for ejection, and the opposite is true when $m_3 > m_1$.  In both limits, the preference for ejection of the lightest object is substantially stronger in escapes and weaker in excursions.  For \(m_{\rm 3} = m_{\rm 1}\) we are in the equal mass case and obtain \(1/3\) for all conditional branching ratios, as symmetry demands. The equal-mass limit also seems to be the turning point for escape vs excursion probabilities. 

Having computed the conditional probability for a given particle to be ejected {\it given} an intermediate or a hyperbolic outcome state, we now compute the branching ratios between these two different types of ejections.  Employing the same ergodic reasoning to produce branching ratios, the probability of a 3-body system disintegrating into a terminal escape at the end of a scramble is
\begin{align}
 p_{\rm dis} = \frac{\sum_{\left(i,j,k\right) \in \{\left(1,2,3\right), \left(2,3,1\right), \left(1,3,2\right)\}} {\sigma_{\rm es}^{i,j,k}}}{\sum_{\left(i,j,k\right) \in \{\left(1,2,3\right), \left(2,3,1\right), \left(1,3,2\right)\}} {\left(\sigma_{\rm es}^{i,j,k} +  \sigma_{\rm ex}^{i,j,k}\right)}}.
 \label{eq:Pdis}
\end{align}
We plot Eq. \ref{eq:Pdis} in Fig. \ref{fig:DisintegrationProb} varying the triplet mass configuration in the same way as in Fig. \ref{fig:EscapeExcursionProb}. We see that the lowest \(p_{\rm dis}\) is obtained in the equal mass case regardless of the total angular momentum and choice of chaotic region criteria. We also see higher disintegration probabilities for lower total angular momentum, suggesting that systems with numerous scrambles are most likely to be ones with high total angular momentum; however, this effect is relatively weak. 

If we treat each scramble as being independent from the last, with each scramble having an identical probability of \(p_{\rm dis}\) to disintegrate, then the probability that the system will undergo $N$ scrambles before disintegrating is a simple geometric distribution, 
\begin{align}
    P(N_{\rm scram} = N) =  (1-p_{\rm dis})^{N-1}p_{\rm dis}.
    \label{eq:PofNscram}
\end{align}
According to numerical three-body simulations done in \citealt{StoneLeigh19}, the assumption of scramble independence is an oversimplification. From that work's analysis, the first two scrambles especially seem to show non-uniformity in the outcome phase space, indicating some memory is retained of initial conditions over a period of 1-2 scrambles. As the system exhibits more scrambles, however, these non-uniform structures dissolve, making the above approximation more reliable for long-lived triples. In a different numerical study done in \citealt{Trani+2024}, it was shown that the non-chaotic regular trajectories can occupy a significant part of phase space, which is sensitive to the initial configuration of the three bodies.
Nonetheless, since taking the phase space non-uniformity into account requires either (i) a detailed model for either islands of regularity in the 3-body problem's phase space or (ii) failures of ergodicity, neither of which has been developed analytically, we retain the assumption of scramble independence even for the first scrambles. 

In Fig. \ref{fig:RandAEvsNumerical} we plot $P(N_{\rm scram}=N)$ from Eq. \ref{eq:PofNscram} against \(N\), using the per-scramble disintegration probability \(p_{\rm dis}\) obtained from integrating Eq. \ref{eq:3ddist} as specified in Eq. \ref{eq:Pdis}.  Against this analytic prediction, we also show $P(N_{\rm scram})$ probabilities obtained from direct integration of the equations of motion in binary-single scattering experiments.  This numerical data is taken from the 3-body simulations of \citet{StoneLeigh19}, for an equal mass distribution \(m_1 = m_2 = m_3 = \Msun\), total energy \(E_0 = - {Gm_1 m_2}/{\rm 1 AU}\) and total angular momentum \(L_0 = \mathcal{M} \sqrt{G m_B {\rm AU} (1-e_0^2)} \).  
Further details about numerical methods are described in \citet{StoneLeigh19}.  We see that for higher value of angular momentum $L_0$ (i.e. lower values of $e_0$), our analytic predictions for $P(N_{\rm scram}=N)$ work quite well for $N>2$.  The analytic predictions become modestly worse for $e_0 = 0.9$, though they are still reliable to within a few $10$s of $\%$.  Inaccuracies rise to the factor $\approx 2$ level for the extreme case of $e_0 = 0.99$. 
Taken as a whole, this comparison suggests that it takes more scrambles for a low angular momentum system to ergodicize across the available phase space. 

The geometric distribution in Eq. \ref{eq:PofNscram} predicts a mean of
$\left\langle N_{\rm scram}\right\rangle \simeq 5\text{--}7$
for the equal-mass configurations considered in Fig. \ref{fig:DisintegrationProb}, with the mean
determined by the disintegration probability $p_{\rm dis}$.
This is smaller than the mean intermediate state count
$N_{\rm IMS}\simeq 20$ adopted by \citet{Samsing18}, which was
calibrated using isotropic three-body scattering experiments.
However, our prediction is broadly comparable to the counts
suggested by the Newtonian cumulative distribution in the upper
panel of Fig.~6 of \citet{Samsing14}, obtained for initially
very hard, circular binaries interacting with an equal-mass single.
A direct quantitative comparison requires care, since the criteria
used to distinguish temporarily hierarchical configurations from
chaotic scrambles differ between the two approaches.
Differences in these criteria and in the sampled initial conditions,
together with possible departures from scramble independence and
ergodicity discussed above, may therefore contribute to the differing
mean counts. For unequal mass configurations, the generally higher disintegration
probabilities (Fig.  \ref{fig:DisintegrationProb}) imply still smaller values of
$\left\langle N_{\rm scram}\right\rangle$, as illustrated by our
simulations in Section~\ref{sec:results}. We further explore the dependence of the predicted expected scramble count on the total
angular momentum and mass ratios in
Fig.~\ref{fig:mean_scrambles_L0} in Appendix \ref{app:beta}.

While Fig. \ref{fig:RandAEvsNumerical} is highly insensitive to our qualitative choice of models for the strong interaction region (i.e. SE versus AE), it is quite sensitive to numerical choices for $\alpha$ (characterizing the physical size of the strong interaction region) and $\beta$ (characterizing the transition between scrambles and ejections).  We explore this dependence in greater detail in Appendix \ref{app:beta} and especially in Figs. \ref{fig:alphasvsNumerical} and \ref{fig:bettasvsNumerical}.  Our primary conclusions here are that (i) as $\alpha$ increases, the $P(N_{\rm scram})$ curve steepens; (ii) as the ratio of $\beta/\alpha$ increases, the $P(N_{\rm scram})$ curve steepens; the best agreement with numerical scattering experiments is achieved for $\beta=\alpha$ when $\alpha \approx 2.5$ for SE and $\alpha \approx 3.5$ for AE.  At a practical level, this suggests that intermediate states can be analytically modeled with a single fudge factor ($\alpha$) and that there is no need to introduce a second, independent one ($\beta$).  Conceptually, this result suggests a hard separation between ejective excursions (the formation of a decoupled, temporarily hierarchical triple) and scrambles, rather than a continuous transition between these two regimes.  We speculate that further investigation of the strong interaction region may require explicit use of a triple stability criterion, but that goes beyond the scope of this work. 

\section{Sampling from Cluster Distributions}
\label{sec:cluster_dist}

Outcomes of non-hierarchical triple evolution may become important in many areas of astrophysics, but one application of recent interest concerns the mergers of stellar mass black holes (BHs), which can be observed via their GW emission.  Repeated binary-single scatterings have long been viewed as a promising astrophysical pathway to binary BH merger \citep{portegies_zwart_black_2000}, and are likely responsible for at least a substantial fraction of the detected binary BH merger population \citep{Rodriguez+16, Kremer+20}.  While most BH mergers catalyzed by binary-single scatterings will occur long after the metastable triple has disintegrated (i.e. in a hyperbolic outcome state), a non-trivial minority will occur during intermediate (elliptic) states of triple evolution, due to close passages that trigger massive GW energy loss and quick inspirals \citep{Samsing18}.  This type of ``prompt'' GW inspiral can make up ~5-10\% \citep{samsing_mocca-survey_2018} of all mergers produced by iterated binary-single scatterings, and carries a relatively unique signature: detectable eccentricity in the LIGO band.  While prompt captures during intermediate states are not the exclusive source of eccentric GW signals, other proposals for achieving high eccentricity inspirals are likely to be intrinsically rare \citep{OLeary+09, Antonini+17}.

In this section, we develop a simple but flexible approach for estimating prompt GW capture rates in dense stellar environments.  In order to calculate merger rates, we implement a Monte Carlo sampling algorithm for the system. 
The following sections will describe the model used for the cluster, the way we sample triples, and how the cluster model parameters affect these samples. The overall philosophy of this section is similar to other works that designed fast binary BH population synthesis techniques \citep{AntoniniGieles20, Mapelli+22, Kritos+24}.  In contrast to these works, the main novelty here is the incorporation and usage of more detailed intermediate state physics developed in \S \ref{sec:IMS} and \S \ref{sec:lifetimes}.

\subsection{Two Component Clusters}
Most stars and binaries in the Universe live in the field, meaning the collisionless environments of galactic disks and bulges.  A minority, however, live in the collisional environments of star clusters, where stellar densities exceed field densities by orders of magnitude, and the mean-free-path for binary-single scatterings is relatively short.  Dense star clusters can be categorized by their origins and structural parameters into several qualitative types: open clusters are the least dense and least tightly bound; globular clusters (GCs) and young massive clusters have greater densities and velocity dispersions; nuclear star clusters (NSCs) are the densest and most tightly bound.  All of these systems are quasi-spherical and in most cases have collisional relaxation times less than their ages, so they will relax into a configuration close to an isothermal sphere (although collective effects may modify the secular relaxation process and the resulting cluster structure; \citealt{Hamilton_2018, Fouvry_2021}).  The exception to this rule is in dense star clusters containing a central massive black hole, which may be the case in some GCs \citep{Haberle+24} and is certainly common in larger NSCs \citep{Neumayer+20}.  The presence of a massive black hole will, however, generate high velocity dispersions that efficiently ionize binaries \citep{heggie_binary_1975}, reducing the suitability of such a system for producing binary BHs from binary-single scatterings.  For the remainder of this paper, therefore, we only consider clusters lacking a massive black hole.

In principle, all four types of clusters enumerated above may contribute a major fraction of the LVK-band GW production rate.  Globular clusters have received by far the most theoretical attention, in part due to their evidenced ability to shuffle compact objects into binaries via binary-single exchanges \citep{Katz75}, and theoretical modeling of their evolution indicates that they may produce a majority of observed binary BH mergers \citep{Rodriguez+16, Askar+17}.   
Open and young massive clusters have received less study, but may also be able to assemble binary BHs at a rate important for GW observatories \citep{FragioneBannerjee22, Kumamoto+20}.  In this section, however, we focus our attention on NSCs.  Their GW production rates are uncertain but potentially important for the LVK sample of binary BHs \citep{AntoniniRasio16}, and their mergers may contain relatively unique signatures.  For example, the high escape velocities of NSCs may help to foster multiple generations of binary BH merger \citep{Mahapatra+21}, hierarchically assembling more massive BHs that can be seen by the LVK interferometers in the ``pair instability gap'' forbidden to single-star evolution \citep{Farmer+19}.  NSCs are also well-suited for study with approximate analytic models.  As the most massive collisional clusters in the Universe, NSCs are (i) more capable of satisfying the statistical limit in their cores, and conversely, are (ii) particularly hard to model using direct $N$-body simulations \citep{Wang+20}.  

One can find a variety of analytic or semi-analytic cluster models across the literature, many of which are summarized in \citet{galacticdynamics2ed}. In this paper, however, we use a three-parameter potential-density pair as presented in \citet{StoneOstriker2015} to model the NSC environments hosting 3-body interactions. We opted for this model due to its analytical tractability, enabling us to readily express essential parameters such as density, potential, and binding energy in closed forms.

The stellar density profile \(\rho_{\rm \circ}\left(r\right)\) employed is the following:
\begin{align}
    \rho_{\rm \circ}\left(r\right) = \frac{\rho_{\rm c}}{\left(1 + r^2/r_{\rm c}^2\right)\left(1 + r^2/r_{\rm h}^2\right)}
    \label{eq:clusterdenstiyprof}
\end{align}
where \(r\) is the spherical radius coordinate, \(\rho_{\rm c}\) is the cluster core density, \(r_{\rm c}\) the core radius and \(r_{\rm h}\) the half mass radius. 
This density profile is then used to calculate the potential \(\Phi_{\rm \circ}\left(r\right)\), the total enclosed stellar mass \(M_{\rm \circ}\left(r\right)\), and the three-dimensional velocity dispersion \(\sigma^2_{\rm \circ}\left(r\right)\).
All of these quantities have explicit analytical radial profiles given in \citep{StoneOstriker2015}, containing only \(r_{\rm h}, r_{\rm c}\) and \(\rho_{\rm c}\) as free parameters.  In the constant-density cluster core, the potential and velocity dispersion asymptote to:
\begin{align}
    \Phi_{\rm \circ}\left(r\rightarrow0\right) = -\frac{1}{\pi}\frac{GM_{\rm tot}}{r_{\rm h}-r_{\rm c}}\ln\left(\frac{r_{\rm h}}{r_{\rm c}}\right),
    \label{eq:clusterpotential}
\end{align}
\begin{align}
    \sigma^2_{\rm \circ}\left(r\rightarrow0\right) =12\pi G\rho_{c}r_{c}^{2}\left(\frac{\pi^{2}}{8}-1\right)
    \label{eq:clusterdispprof},
\end{align}
respectively.  The cluster escape velocity will often be important for estimating when binary-single scatterings terminate, and can be computed from the potential as $v_{\rm esc}(r) = \sqrt{-2\Phi_\circ(r)}$.  Conversely, the total cluster mass is given by
\begin{align}
    M_{\rm \circ}\left(r\rightarrow\infty\right) = \frac{2\pi^{2}r_{\rm c}^{2}r_{\rm h}^{2}\rho_{\rm c}}{r_{\rm h}+r_{\rm c}} ,
    \label{eq:clustermassprof}
\end{align}
which for some cases is a more useful third parameter than $\rho_{\rm c}$.  Likewise, it is more convenient to use the unit-less cluster concentration parameter \(c_{\rm \circ} = r_{\rm h}/r_{\rm c}\) instead of \(r_{\rm c}\).

Many simulations of dense cluster dynamics find the formation of a central BH sub-cluster (\citealt{banerjee_stellar-mass_2010}, \citealt{downing_compact_2011}, \citealt{breen_dynamical_2013}, \citealt{sedda_dragon-ii_2023}, \citealt{arca_sedda_mocca-survey_2018}, \citealt{kremer_role_2019}). This BH sub-cluster can form {\it in place} if massive stars sink to the cluster center before evolving to BHs, although in extreme cases, this situation can also produce an intermediate-mass BH through runaway collisions \citep{portegies_zwart_runaway_2002,portegies_zwart_formation_2004,rizzuto_intermediate_2021}.  Alternatively, since the lifetime of massive stars is short, they may become stellar mass BHs at larger radii in the cluster and then only sink to the center on a dynamical friction timescale.   

Once substantial numbers of BHs begin arriving at the center of the overall star cluster, they may assemble a long-lived sub-cluster, as was suggested by older analytic arguments \citep{spitzer_equipartition_1969}.  However, the BHs may also efficiently self-eject, either entirely from the cluster or into the lower-density cluster halo, preventing the assembly of a central, BH-dominated sub-cluster \citep{morscher_retention_2013, morscher_dynamical_2015}.  Long-lived BH sub-clusters are more common for higher mass star clusters \citep{sedda_dragon-ii_2023}.

These results motivate us to use a two-component cluster, where a BH sub-cluster (indicated with a subscript of \(\rm BH\)) and a parent star cluster (indicated with a subscript of \(\star\) ) are both described independently by Eq. \ref{eq:clusterdenstiyprof}. The total, two-component cluster (indicated with a subscript of \(\rm \varocircle\)) attains a density which is a linear sum of BH and star cluster density profiles:
\begin{align}
    \rho_{\varocircle}\left(r\right) = \rho_{\rm BH}\left(r\right) + \rho_{\rm \star}\left(r\right).
    \label{eq:twocompdensity}
\end{align}
The potential and total cluster mass are added linearly, while the escape velocity and the three-dimensional dispersion attain a more complicated relation to the density profiles. These properties are calculated by plugging in Eq. \ref{eq:twocompdensity} into Eqs. \ref{eq:clusterpotential}-\ref{eq:clusterdispprof}.  Under the assumption of rapid mass segregation and our interest in BH binaries, we will restrict further discussion to the cluster core values only \(r \ll r_{\rm c}\).

To calibrate the free parameters of this model, we assume that the stars and BHs are equilibrated to equal dispersion velocities (an ``isovelocity'' rather than isothermal state).  While this assumption violates the principle of equipartition \citep{spitzer_equipartition_1969}, modern N-body simulations of multi-species dense clusters generally find large deviations from equipartition that are generally closer to the isovelocity limit \citep{arca_sedda_mocca-survey_2018,trenti_no_2013,webb_link_2017}. Using the analytical formula for the dispersion velocity at the cluster core (Eq. 11 in \citealt{StoneKupperOstriker2017}) twice to equate the dispersion velocities, we arrive at the relation \(r_{\rm h, BH} = \left(M_{\rm tot, BH} / M_{\rm tot, \star}\right) r_{\rm h, \star}\).

 The total BH cluster mass \(M_{\rm tot, BH}\) is normalized by employing the ratio of the number of BHs \(N_{\rm BH}\) to that of stars \(N_{\rm \star}\), \(f_{\rm BH} = N_{\rm BH}/N_{\rm \star}\).  We compute this as \(M_{\rm tot, BH} = f_{\rm BH} N_{\rm \star} m_{\rm BH}\) where \(N_{\rm \star} = M_{\rm tot, \star}/m_{\rm \star}\), \(m_{\rm BH}\) is the average BH mass and \(m_{\rm \star}\) is the average star mass.
We likewise employ the BH concentration parameter \(c_{\rm BH} = r_{\rm h, BH}/r_{\rm c, BH}\) as a proxy for the BH sub-cluster core radius. 
Throughout the Monte Carlo calculations, we use the values \(f_{\rm BH} \simeq 10^{-3}\), \(c_{\rm BH} \simeq 3.5\) in accordance with \citep{sedda_dragon-ii_2023}, and take \(m_{\rm \star} = 0.38 M_{\rm \varodot}\), \(m_{\rm BH} = 20 M_{\rm \varodot}\).

With these assumptions, each point in the 3d parameter space of \(\left(M_{\rm tot, \star}, r_{\rm h, \star}, r_{\rm c, \star}\right)\) defines a two-component cluster. Observations of NSCs \citep{GeorgievBoker14} find significant correlations between these three parameters \citep{StoneKupperOstriker2017}, so it is sufficient to explore a limited portion of this parameter space to gain insights into the observed set of GCs. 
We fit a single power law to the \(M_{\rm tot, \star}-\sigma_{\rm \star}\) data presented in \citep{StoneKupperOstriker2017} Fig. 2, we find:
\begin{equation}
    M_{\rm tot, \star} = M_{\rm l}\left(\frac{\sigma_{\rm \star}}{\rm km / s}\right)^{s},
    \label{eq:MtotOfSigma}
\end{equation}
where \(M_{\rm l} \simeq 10^{3.42} M_{\rm \varodot}, s \simeq 1.86\) are the fitting parameters. 
We then expand Eq. \ref{eq:clusterdispprof} to leading order in $r_c/r_h$ and compute
\begin{equation}
    r_{\rm h, \star}\left(M_{\rm tot, \star}\right) = \frac{6G\left(\frac{\pi^2}{8} - 1\right) M_{\rm l}^{\frac{2}{s}}}{\pi M_{\rm tot, \star}^{\frac{2}{s} - 1} \left(\frac{\rm km}{s}\right)^2}.
    \label{eq:rhOfMtot}
\end{equation}
Throughout the simulations, we explore clusters in this 2d parameter space in two ways.  First, we explore \(M_{\rm tot, \star} \in \left[10^{5} M_{\rm \varodot}, 10^{8} M_{\rm \varodot}\right]\) with a fixed \(c_{\rm \star} \simeq 6.3\). We then choose a fixed \(M_{\rm tot, \star} \simeq 10^{6.33} M_{\rm \varodot}\) and explore the stellar concentration parameter \(c_{\rm \star} \in \left[2, 20\right]\).

\subsection{Sampling Incoming Singles}
To construct a chaotic, non-hierarchical triple, first consider a pre-existing binary in a dense star cluster. Given a binary with masses \(m_{\rm a}, m_{\rm b}\), energy \(E_{\rm B}\), angular momentum \(L_{B}\) and cosine-inclination \(C_{Bd}\) (this inclination variable is defined with respect to an external axis) inside a cluster with dispersion velocity \(\sigma_{\rm \circ}\), we must sample an incoming single from the cluster.  
We sample the speed of the incoming single \(v_{\rm s}\) from a Maxwell-Boltzmann distribution
\begin{align}
    f_{v_{s}}\left(v_{\rm s}\right)\propto v_{\rm s}^2\exp\left({{-\frac{v_{\rm s}^2}{2\sigma_{\rm \circ}^2}}}\right).
    \label{eq:NormalSpeedDist}
\end{align}
This sets the total energy of the resulting triple, \(E_{\rm 0} = E_{\rm B} + \frac{1}{2}mv_{\rm s}^2\). If the total energy is positive but only slightly so, with \(\sqrt{|E_{\rm s}/E_{\rm B}|} < 1.5\), the triple is considered to undergo an \textit{exchange}. This is in accordance with results found in three body numerical studies of \citealt{HutBahcall1983} and \citealt{Fregeau+2004}.  We approximate this by choosing the two most massive objects and keeping the energy and eccentricity of the original binary. 
If the total energy is positive and \(\sqrt{|E_{\rm s}/E_{\rm B}|} > 1.5\), all the bodies in the triple are considered to be unbound and scatter in different directions and the simulation ends with an \textit{ionization} (i.e. total dissolution). If the total energy is negative, a chaotic, non-hierarchical triple is considered to form.  

If a non-hierarchical triple forms, we then sample the cosine-inclination of the tertiary's orbit with respect to the cluster axis, \(C_{s}\), from a uniform distribution in the range \(\left[-1,1\right]\) and calculate the mutual inclination \(C_{\rm B}\) from vector arithmetic, 
\begin{align}
    C_{\rm B} = C_{\rm s}C_{\rm Bd}\pm \sqrt{\left(1 - C_{\rm Bd}^2\right)\left(1 - C_{\rm s}^2\right)}.
    \label{eq:CBscalc}
\end{align}
The ambiguous sign is determined by the sign of \(\sin{I_{\rm Bd}\sin{I_{\rm s}}}\). Since the inclination \(I_{\rm s}\) distribution is symmetric, this sign is chosen randomly by a fair coin flip. 
We can now proceed to sample the impact parameter \(b\) and in turn, calculate the tertiary's angular momentum \(L_{\rm s} = mbv_{\rm s}\).  

Non-hierarchical triples cannot form for arbitrarily large $b$; equivalently, there is an upper bound on \(L_{\rm 0}\) for such a system, denoted as \(L_{\rm 0, max}\) and derived in Appendix \ref{app:L0maxder}.  By angular momentum conservation, upper and lower bounds on \(L_{\rm s}\), denoted by \(L_{\rm, s \pm}\), must exist; this relation can be turned around to place bounds on the impact parameter 
\begin{align}
    \frac{L_{\rm s, -}}{mv_{\rm s}} < b < \frac{L_{\rm s, +}}{mv_{\rm s}}
    \label{eq:bLscalc}
\end{align}
where 
\begin{align}
    L_{\rm s, \pm} = -L_{\rm B} C_{\rm B} \pm \sqrt{L_{\rm 0, max}^2 - L_{\rm B}^2 \left(1 - C_{\rm B}^2\right)}
    \label{eq:Lsplus}
\end{align}

To derive the distribution of impact parameters, we consider a binary in a background of single stars. The binary-single interaction rate for encounters with some cross-section $\Sigma$ will be given by $\dot{N} = n \Sigma v_{\rm s}$, where $n$ is the number density of scatterers, and  
\begin{align}
    \Sigma = \frac{2\pi R G M}{v_{s}^2},
    \label{eq:SigmaFocused}
\end{align}
where $R$ is the same approximation to the chaotic triple boundary previously introduced in \S \ref{sec:derivation}.  Combining with Eq. \ref{eq:bLscalc}, we arrive to the maximum permitted impact parameter \(b_{\rm max}\), defined as
\begin{align}
    b_{\rm max} = \min \left\{\frac{\sqrt{2RGM}}{v_{s}}, \frac{L_{\rm s, +}}{mv_{\rm s}}\right\}.
    \label{eq:impactParamMax}
\end{align}
The probability distribution function for encountering a single star with impact parameter \(b\) and velocity \(v_{s}\)is simply
\begin{align}
    \frac{d\dot{N}}{db} = 2 v_{s} n \pi b.
    \label{eq:impactParamDist}
\end{align}
Therefore, we sample the impact parameter of the tertiary \(b\) from a linear distribution \(f_{\rm b}\left(b\right)\propto b\) allowed between 0 and \(b_{\rm max}\left(v_{\rm s}\right)\).
We can now construct the total angular momentum of the triple, \(L_{\rm 0}\), and also the cosine-inclination of the angle between the total angular momentum direction and the cluster axis:
\begin{align}
    C_{\rm 0} = \frac{L_{\rm B}C_{\rm Bd} + L_{\rm s}C_{\rm s}}{L_{\rm 0}}.
\end{align}
With the triple properties in hand, we can use Eq. \ref{eq:3ddist} here and Eq. 17 in \citet{StoneLeigh19} to sample binary properties given an intermediate state (IMS) or a final state (FS), respectively. Going further, we can sample the number of IMSs of the triple, \(N_{\rm scram}\), from Eq. \ref{eq:PofNscram}.  We use the SE chaotic region criteria to construct all three-body distributions. We ultimately sample from the distribution of intermediate states a total of \(N_{\rm scram}\) times, and from the FS distribution one time, to construct a sequence of binary properties during times of dynamical isolation from the tertiary. We call this sequence \textit{the binary evolution}.

This set of prescriptions is crude, as it neglects or oversimplifies various dynamical features of binary-single scattering:
\begin{itemize}
    \item Our model for the two-component cluster is approximate; importantly, it neglects time evolution (in the form of binary burning and gravothermal oscillations) and realistic mass sepctra.
    \item Encounters with very large $b$, which fall into the secular regime \citep{HamersSamsing19, leigh_thermodynamics_2022}, are completely neglected.
    \item Strong encounters that fail to produce a meta-stable, non-hierarchical triple (i.e. those that resolve as a non-chaotic flyby or prompt exchange) are also generally neglected.
\end{itemize}
We will return to the limitations of this approach in \S \ref{sec:conclusions}.

\subsection{Initial Binary}
Although the sampling algorithm above can be applied to any particular binary, we are interested in estimating features of GW captures, so we restrict ourselves to BBHs. Since most BBHs are assembled from main sequence (MS) stellar binaries that underwent dynamical swaps with single BHs \citep{morscher_retention_2013}, we build our BBH initial conditions by sampling from a population of MS binaries and introducing single BHs.

The MS binary semi-major axis \(a_{\rm B}\) is sampled from an Opik's law distribution (\(\propto 1/a\) ), bounded from above by the semi-major axis for tidal disruption in the cluster core, \(a_{\rm t} = \left(m_{\rm B}/\rho_{\rm c}\right)^{1/3}\), and from below by the semi-major axis for a physical collision, \(a_{\rm coll} = R_{\rm \star}\left(m_{\rm a}\right) + R_{\rm \star}\left(m_{\rm b}\right)\), where the stellar radius \(R_{\rm \star}\) is given by a standard mass-radius relation \citep{kippenhahn_stellar_1990}:
\begin{align}
    R_{\rm \star} =
        \begin{cases}
            R_{\rm \varodot}\left(\frac{m_{\rm \star}}{M_{\rm \varodot}}\right)^{0.8} & m_{\rm \star} \leq M_{\rm \varodot} \\
            R_{\rm \varodot}\left(\frac{m_{\rm \star}}{M_{\rm \varodot}}\right)^{0.55} & m_{\rm \star} > M_{\rm \varodot}                      .
    \end{cases}
    \label{eq:starmass}
\end{align}
Here \(m_{\rm \star}\) is the stellar mass and \(R_{\rm \varodot}, M_{\rm \varodot}\) are, respectively, the Solar radius and Solar mass.  Individual masses $m_{\rm a}$ and $m_{\rm b}$ are taken from a Kroupa initial mass function as will be described below.  
The initial binary eccentricity \(e_{\rm B}\) is then sampled from a thermal distribution (\(\propto 2e_{\rm B}\)) from 0 to \(1 - a_{\rm coll}/a_{\rm B}\), in rough agreement with field eccentricity distributions \citep{MoeDiStefano2017}, but tuned to prevent immediate collisions.
The cosine-inclination of the binary is sampled isotropically, i.e. uniformly within the range of \(\left[-1, 1\right]\).

We introduce a single BH and use Eq. \ref{eq:Pm3es} three times (each time permuting the three masses) to calculate the probabilities of different binaries surviving the chaotic interaction given an escape. We probabilistically sample which of the binaries survives according to these branching ratios, and compute the FS outcome in the manner described in the previous section.  We repeat this procedure, repeatedly re-introducing a single BH, until a BBH is formed. We call this process the \textit{pre-evolution} of the binary.
If a constructed binary in the pre-evolution stage finds its end in a direct collision or stellar tidal disruption, then it is discarded and a new binary is pre-evolved.

\subsection{Mass Distribution Models within Clusters}
We consider three different models for the mass distributions within NSCs.  The first and simplest among them is the {\it equal mass BBH} case, where the pre-evolution is done with stars of mass \(m_{\rm \star} = 0.38 M_{\rm \varodot}\) and all BHs are of mass \(m_{\rm BH} = 20 M_{\varodot}\). The astrophysical motivations for this case are (i) mass segregation, which makes similar mass objects more likely to interact with each other \citep{weatherford_black_2021}, and (ii) the larger gravitationally focused cross-section for BHs as opposed to main sequence stars (Eq. \ref{eq:SigmaFocused}).  Another motivation is computational simplicity, for each IMS or FS one does not have to calculate the identity of the tertiary as all 3 masses are the same. This reduces the number of computed integrals by at least a factor of 3 when calculating the disintegration probabilities.

Next we consider the {\it unequal mass} case, where different masses of stellar and compact objects within the cluster are accounted for. A present day mass function (PDMF) is constructed from the Kroupa initial mass function (IMF) \({\rm d}N/{\rm d}M_{\rm ZAMS}\) which describes the distribution of zero-age main sequence (ZAMS) masses \(M_{\rm ZAMS}\):
\begin{align}
    \frac{{\rm d}N}{{\rm d}M_{\rm ZAMS}} \propto 
        \begin{cases}
            M_{\rm ZAMS}^{-1.3} & 0.08 \leq M_{\rm ZAMS}/M_{\rm \varodot} < 0.5 \\
            M_{\rm ZAMS}^{-2.3} & 0.5 \leq M_{\rm ZAMS}/M_{\rm \varodot} < 1 \\
            M_{\rm ZAMS}^{-2.7} & 1 \leq M_{\rm ZAMS}/M_{\rm \varodot},                      
        \end{cases}
\end{align}
along with the initial-final mass relationship \citep{webb_back_2015}.  Here we assume that all stars with zero-age main sequence mass (ZAMS) that are below \(M_{\rm ZAMS} < 8 M_{\rm \varodot}\) end up as \(0.6 M_{\rm \varodot}\) white dwarfs (WD), all ZAMS masses \(8\leq M_{\rm ZAMS}/M_{\rm \varodot} \leq 25 \) end up as \(1.4 M_{\rm \varodot}\) neutron stars (NS), all ZAMS masses \(25\leq M_{\rm ZAMS}/M_{\rm \varodot} \leq 110 \) and \(230\leq M_{\rm ZAMS}/M_{\rm \varodot}\) end up as BHs, while the intermediate range \(110< M_{\rm ZAMS}/M_{\rm \varodot} < 230\) is completely destroyed in pair instability supernovae.
The BH final mass is taken from a fitting formula to the \citet{spera_very_2017} Table D3 data for the low metallicity \(Z = 2 \times 10^{-4}\) clusters, which approximates the complicated relationship between the progenitor star mass and the remnant BH mass\footnote{We consider low-metallicity NSCs for specificity here; in principle, we could have considered higher metallicity systems without qualitatively changing our results.  The main difference would be the disappearance of the pair instability mass gap, and the suppression of high BH birth masses \citep{spera_very_2017}.}.

The initial binary masses are sampled from the constructed main sequence mass distribution and all singles are sampled from the total constructed PDMF.  We note that this procedure does not account for changes to the cluster core PDMF due to mass segregation among the main sequence stars \citep[e.g. ][]{leigh_quantifying_2012}, but we expect the impact of this to be relatively modest in systems where the core is dominated by a BH population, as are of the greatest interest for us.

Lastly, we have the {\it unequal mass BBH} case, where we pre-evolve an initial MS binary sampled from a MS PDMF constructed as before, and sample single BHs from the BH PDMF until a binary BH is formed. We then sample the next single object from the whole PDMF without restraining it to BHs or any other stellar object. This case zooms in on the dynamically formed BBH inside the NSC, and provides sequences for their evolution only. Since most GW inspirals observed today are thought to be of BH-BH origin, this is the case of greatest interest to us.

\subsection{Binary Properties for Physical Dissolution}
\label{sec:binary_props_for_diss}
During the binary evolution, the IMS and FS binaries may be destroyed through a number of mechanisms, namely ionizations, collisions, mergers and tidal disruption events (TDEs).  Our treatment of ionizations has already been mentioned above.  Mergers and TDEs may occur when IMS or FS binaries are created with very high eccentricities, such that the binary pericenter falls below a critical threshold. 

In the case of a binary containing MS stars or a WD and a MS star, with a sufficiently small pericenter, a \textit{collision} will take place. This pericenter is taken to be the maximal radius between the radii of the binary objects. To compute a MS star radius \(R_{\rm \star}\), we use Eq. \ref{eq:starmass}.
For the WD radius \(R_{\rm WD}\), we use the approximate relation \citep{kippenhahn_stellar_1990}:
\begin{align}
    R_{\rm WD} = 4.2\times 10^{-3}  \left(\frac{M_{\rm \varodot}}{m_{\rm WD}}\right)^{\frac{1}{3}} R_{\rm \varodot}.
    \label{eq:WDradius}
\end{align}
Following creation of an IMS or FS binary, we first check if a collision has occurred; if yes, we terminate the evolution, but if no, we next examine whether a \textit{TDE} has happened. We consider that a TDE can occur as long as at least one binary component is either a MS or a WD star.  If the binary pericenter \(q_{\rm B}\) is less than the tidal radius \(r_{\rm TDE}\), a TDE is considered to take place. Here the tidal radius is given by \citep{rees_tidal_1988}:
\begin{align}
    r_{\rm TDE} = r_{\rm 2}\left(\frac{m_{\rm 1}}{m_{\rm 2}}\right)^\frac{1}{3},
    \label{eq:TDEradius}
\end{align}
where \(m_{\rm 2} < m_{\rm 1}\) are the binary masses.

In this work we neglect an additional source of eccentricity excitation, namely perturbations from weak and distant tertiary flybys, as we are primarily focused on hard binaries where these encounters are less important \citep{reinoso_mean_2022}.  However, we note that a succession of such weak encounters may deplete the high-eccentricity tail of the binary population, and in principle these effects can be accounted for in a semi-analytic formalism such as ours \citep{leigh_thermodynamics_2022}, though we leave this for future work.

\subsection{CO Mergers: Eccentric and Non-Eccentric}
In the case of a CO binary with a sufficiently small separation, a \textit{merger} may occur. If the eccentricity exceeds a detectable threshold, the merger can leave a distinctive signature in the GW signal, and will be considered an \textit{observably eccentric merger} (OEM).

To calculate the limiting binary properties for mergers, we make use of the largest pericenter available for an IMS BBH merger with a detectable eccentricity, \(q_{\rm B}^{OEM}\), as derived in \citep{Samsing18},
\begin{align}
    q_{\rm B}^{\rm OEM} \equiv q_f \frac{1}{2F\left(e_f\right)}\left(\frac{425}{304}\right)^{\frac{870}{2299}},
    \label{eq:qBEMIMS}
\end{align}
where \(q_f\) is the pericenter distance for a BBH to emit GWs with peak frequency \(f\),
\begin{align}
    q_f \equiv \left(\frac{G m_{\rm B}}{\pi^2 f^2}\right)^\frac{1}{3},
    \label{eq:qf}
\end{align}
\(e_f\) is the eccentricity at frequency \(f\), and \(F\left(e\right)\) is the dimensionless function
\begin{align}
F\left(e\right) = \frac{e^{12/19}}{1+e}\left(1+\frac{121}{304}e^2\right)^{\frac{870}{2299}}.
\end{align}
We take \(e_f = 0.1\) and \(f = 10  {\rm Hz}\) in the subsequent calculations as a rough proxy for minimum robustly detectable eccentricities with the current LVK network, but note that this is an approximate threshold and may change for individual events, or for future improvements in GW detectors.

For the intermediate state binary to merge before the tertiary comes back from its excursion, the timescale for GW erosion of the semimajor axis \(T_{\rm GW}\) has to be smaller than the orbital period of the tertiary \(T_{\rm s}\), i.e. we require $T_{\rm GW} < T_{\rm s}$.  
Assuming that almost all prompt mergers come from binaries born with high eccentricities, we can employ the \citet{Peters1964} equation for \(T_{\rm GW}\) in the $e\to 1$ limit. Since the system spends most of the decay time in a state where $a_{\rm B}$ has not yet shrunk dramatically from its initial value, this inequality becomes
\begin{equation}
    T_{\rm GW} \simeq \frac{3}{85} \frac{a_B^4 c^5 \left(1-e_B^2\right)^{\frac{7}{2}}}{G^3 m_{\rm a} m_{\rm b} m_{\rm B}} < 2\pi \sqrt{\frac{a_{\rm s}^3}{Gm_{\rm B}}} = T_s,
    \label{eq:TGWlessTsExplicit}
\end{equation}
with $c$ the speed of light.  Rewriting, we get the following maximal binary pericenter for a GW capture:
\begin{equation}
    q_{\rm B}^{\rm Merge} \equiv \left(\frac{85\pi}{3}\right)^{\frac{2}{7}}\frac{G}{2}\left(\frac{m_B^4 m_s^3 E_B m_a m_b}{\left(E_0 -E_{\rm B}\right)^3 c^{10}}\right)^{\frac{1}{7}}.
    \label{eq:qBMergeIMS}
\end{equation}
Therefore, the pericenter bounds for the IMS phase space volume \(\sigma_{\rm e, merger}^{\rm OEM}\) permitting observably eccentric IMS mergers will be 
\begin{equation}
    q_{\rm B} \in [0,\min\{q_{\rm B}^{\rm OEM}, q_{\rm B}^{\rm Merge}\}].
    \label{eq:qEMmerge}
\end{equation}
Likewise, the bounds for the phase space volume \(\sigma_{\rm e, merger}^{\rm NEM}\) containing all IMS mergers that are {\it not} observably eccentric will be 
\begin{equation}
    q_{\rm B} \in [\min\{q_{\rm B}^{OEM}, q_{\rm B}^{\rm Merge}\}, q_{\rm B}^{\rm Merge}] .
\end{equation}

Similarly, we derive the probability for a gravitational wave capture after formation of a FS binary. In order to find a constraint analogous to Eq. \ref{eq:TGWlessTsExplicit}, we must assume a reference time for the FS merger, \(T_{\rm ref}\). Inside NSCs, the merger must take less time than the typical interval between binary-single scatterings, so \(T_{\rm ref} = T_{\rm bs}\); conversely, while once the system has been ejected from the cluster, the reference time becomes the Hubble time, i.e. \(T_{\rm ref} = t_{\rm H}\).

The binary-single scattering time \(T_{\rm bs}\) is \citep{valtonen_three-body_2006}
\begin{align}
    T_{\rm bs} = \frac{\sigma_{\rm \varocircle}}{2 \pi G a_{\rm B} M n_{\rm \varocircle}\left(r=0\right)},
    \label{eq:binarysinglescatteringtime}
\end{align}
where $\sigma_{\rm \varocircle}$ is the 3d Maxwell-Boltzmann dispersion velocity and \(n_{\rm \varocircle}\) is the total number density given by
\begin{align}
    n_{\rm \varocircle} = n_{\rm \star} + n_{\rm BH} = \frac{\rho_{\rm \star}}{m_{\rm \star}} + \frac{\rho_{\rm BH}}{m_{\rm BH}}
    \label{eq:numberdenisty}
\end{align}

We now derive a criterion for FS mergers using a general reference time \(T_{\rm ref}\), demanding that $T_{\rm GW} < T_{\rm ref}$.  
This inequality plus standard leading-order GW inspiral equations \citep{Peters1964} constrains the energy of the binary
\begin{align}
    \label{eq:hyperEGWbound}
    &E_{\rm B} <  E_B^{\rm GW} \equiv  \\ 
    &-\frac{Gm_{\rm a} m_{\rm b}\left(1-e_{\rm B}^2\right)}{2e_{\rm B}^{\frac{12}{19}}\left(1+\frac{121}{304}e_{\rm B}^2\right)^{\frac{870}{2299}}}
    \left(\frac{12}{19\eta T_{\rm ref}}\int_{\rm 0}^{e_{\rm B}} \frac{{\rm d}e e^{\frac{29}{19}}\left(1+\frac{121}{304}e^2\right)^{\frac{1181}{2299}}}{\left(1-e^2\right)^{\frac{3}{2}}}\right)^{\frac{1}{4}}, \nonumber \notag 
\end{align}
where 
\begin{align}
    \eta = \frac{64}{5}\frac{G^3m_{\rm a} m_{\rm b} m_{\rm B}}{c^5}
    \label{eq:eta}
\end{align}
is used for short.
Using an analytical approximation of \(T_{\rm GW}\)  (labeled \(\tilde{T}_{\rm GW}\); taken from \citealt{Mandel2021}) we can rewrite Eq. \ref{eq:hyperEGWbound} in the following explicit form
\begin{align}
    \label{eq:hyperEGWboundApprox}
    &E_{\rm B} <  \tilde{E}_{\rm B}^{GW} \equiv  \\ 
    &-\frac{Gm_{\rm a} m_{\rm b}}{2}\left(\frac{\left(1+0.27e_{\rm B}^{10}+0.33e_{\rm B}^{20}+0.2e_{\rm B}^{1000}\right)\left(1-e_{\rm B}^2\right)^{\frac{7}{2}}}{4T_{\rm ref} \eta}\right)^{\frac{1}{4}}. \nonumber
\end{align}
Since our constraint for the hyperbolic case is in the \(\{E_{\rm B}, e_{\rm B}\}\) phase space, we integrate Eq. \ref{eq:3ddist} with \(e_{\rm B} \in [0,1], C_{\rm B}\in [-1,1]\) while changing the bounds of \(E_{\rm B}\) to correspond to different types of mergers. For detectably eccentric mergers we rewrite Eq. \ref{eq:qBEMIMS} using the binary eccentricity and energy in the following way:
\begin{align}
    E_{\rm B} <  E_{\rm B}^{OEM} \equiv - \frac{Gm_{\rm a} m_{\rm b} \left(1-e_{\rm B}\right)}{5.4\left(\frac{Gm_{\rm B}}{\pi^2 f^2}\right)^{\frac{1}{3}}}.
    \label{eq:hyperEGWEMboundApprox}
\end{align}
This completes our classification scheme for identifying mergers of COs along a binary evolution track and for separating eccentric from non-eccentric mergers.

\subsection{Repeated Scatterings}
\label{sec:repeated_scatterings}
Now we consider a binary undergoing repeated binary-single scatterings inside a NSC. We shall call this sequence of binary evolutions (which are also sequences of binary states) a dynamical binary sequence (DBS) evolution. 
Assuming no physical dissolution occurs, then with each scattering the resulting binary either hardens, reducing its semi-major axis, or softens, increasing its semi-major axis.

One way these binary-single scatterings can come to an end is when the binary hardens to a point where the kinetic energy it receives in the recoil kick of the FS is enough to be ejected from the NSC. The threshold energy for ejection comes from equating the binary recoil velocity to the cluster central escape velocity, i.e.
\begin{align}
    E_{\rm ej} = \frac{m_{\rm B}}{2}\left(1 + \frac{m_{\rm B}}{m_{\rm s}}\right)v_{\rm esc}^2.
    \label{eq:EjectionEnergy}
\end{align}
This condition is inherently approximate, given the role of the triple center of mass velocity vector in determining cluster escape for any inidividual FS, but it does represent an acceptable average (to be improved upon in future work) over all possible orientations of the center of mass velocity vector.  

The energy of the kick is given by \(E_{\rm 0} - E_{\rm B}\) where \(E_{\rm B}\) is the binary energy in the final state. So, when the following is satisfied:
\begin{align}
    E_{\rm kick} = E_{\rm 0} - E_{\rm B} \geq E_{\rm ej}
    \label{eq:EnergykickForEj}
\end{align}
The binary is considered ejected from the cluster.  
This \textit{ejected binary} state of the binary is not immune to physical dissolutions and an \textit{ejected merger}, \textit{ejected collision}, and \textit{ejected TDE} can all occur if the binary separation conditions are met after ejection. 

Another way a DBS can meet its end is by tidal separation of the binary (TS) by the cluster. If the binary separation is too large, then the differential gravitational pull of the cluster's mass can rip the bodies composing the binary apart. Hence, in each state of the DBS we compare the semi-major axis with the tidal separation semi-major axis, \( a_{\rm TS}=\left(\rho_{\rm c}/m_{\rm B}\right)^{-1/3}\). If $a_{\rm B} > a_{\rm TS}$ during an IMS, the simulation terminates with an \textit{IMS TS} dissolution. If it is bigger during a FS, then the simulation ends with a \textit{FS TS}. If the binary stays intact but (during an IMS) the tertiary is on a large enough excursion for it to be tidally stripped, then a new single is sampled and the evolution continues.

From Heggie's law \citep{heggie_binary_1975} we know that soft binaries tend to continue softening with every binary-single interaction while hard binaries tend to continue hardening. This hard-soft boundary can be represented by the semi-major axis \(a_{\rm HB}\) found from compaing binary binding energy to the average kinetic energy of a single star 
\begin{align}
    a_{\rm HB} = \left(1-\frac{1}{k}\right)\frac{G m_{\rm a} m_{\rm b}}{3\langle m_{\rm \varocircle}\rangle \sigma_{\rm \varocircle}^2}
    \label{eq:HardSofta}
\end{align}
where $\langle m_{\rm \varocircle} \rangle$ is the average mass of an incoming single and $k = E_{\rm 0} / \langle E_{\rm B} \rangle \simeq 3/2$ the ratio of a typical triple's total energy to the expected binary energy in the unbound case.

Given enough time, Heggie's law implies that all DBSs will eventually find their end, though a Hubble time might not be long enough.  If more than a Hubble time passes during the sum of all excursion times within all metastable triples, and the sum of all binary-single scattering times between formation of metastable triples, and no other binary resolution is obtained, then the DBS ends with a binary which is considered an \textit{in-cluster binary}, i.e. a survivor.
We assume the binary evolution time is roughly the sum of the excursion times (as set by the orbital period of the tertiary \(T_{\rm s}\)), and that the times between the last IMS and the next metastable triple can be given by the binary-single scattering time, \(T_{\rm bs}\).

Throughout this work we treat the cluster properties as fixed during each dynamical binary sequence. In reality, two-body relaxation leads to secular evolution of the cluster structure and of the central BH population. We can estimate the importance of this approximation by comparing to the half-mass relaxation time,
\begin{align}
t_{\rm rh}
\simeq
\frac{0.138 N_\star^{1/2}r_{\rm h,\star}^{3/2}}
{\sqrt{Gm_\star}\ln\Lambda},
\end{align}
where we take $\ln\Lambda\simeq\ln(0.4N_\star)$ for an order of magnitude estimate \citep{alexander_gieles_lamers_2014,spitzer_hart_1971}. Using Eq.~\ref{eq:rhOfMtot} and $m_\star=0.38M_\odot$, the cluster models with $M_{\rm tot,\star}=10^5$--$10^8M_\odot$ considered here have $t_{\rm rh}\simeq1.1$--$10.1\,{\rm Gyr}$; for the fiducial $M_{\rm tot,\star}=10^{6.33}M_\odot$ cluster, $t_{\rm rh}\simeq2.9\,{\rm Gyr}$. These values should be regarded as characteristic timescales, since the stellar mass spectrum and central BH subsystem modify the effective relaxation rate.

For comparison, we examined 4000 DBS realizations for a cluster
with $M_{\rm tot,\star}=10^{6}M_{\odot}$ and $c_{\star}=6.3$ and,
after excluding sequences ending as ionized triples or surviving
in-cluster binaries, obtained an estimated mean lifetime of
approximately $6\times10^{7}\,\mathrm{yr}$.
This corresponds to only $\sim0.4\%$ of a Hubble time and is
much shorter than the characteristic cluster evolution timescales
of $\sim1\text{--}10\,\mathrm{Gyr}$ estimated above.

As a simple estimate of the resulting structural evolution, we consider a cluster in balanced evolution, for which the outward energy flux driven by two-body relaxation is commonly parameterized as \citep{alexander_gieles_lamers_2014},
\begin{align}
\dot E = \zeta\frac{|E|}{t_{\rm rh}}, \qquad \zeta\simeq0.1.
\end{align}
Neglecting mass loss and assuming approximately homologous evolution,
$E\propto-M_{\rm cl}^2/r_{\rm h}$. Since
$t_{\rm rh}\propto r_{\rm h}^{3/2}$ at fixed cluster mass, this gives
\begin{align}
\frac{r_{\rm h}\left(t\right)}{r_{\rm h,0}} = 
\left[ 1+\frac{3\zeta}{2} \frac{t}{t_{{\rm rh},0}} \right]^{2/3}.
\end{align}
For the cluster model adopted here, keeping the mass and concentration fixed gives
$
\rho_{\rm c}\propto r_{\rm h}^{-3},
\,
\sigma_{\rm c}\propto r_{\rm h}^{-1/2}$
and hence $T_{\rm bs}\propto\frac{\sigma_{\rm c}}{n_{\rm c}}\propto r_{\rm h}^{5/2}.$
After $1$, $3$, and $5$ initial relaxation times, this simple model therefore predicts increases in $T_{\rm bs}$ by factors of approximately $1.26$, $1.86$, and $2.54$, respectively. 
The corresponding changes in the FS merger threshold are smaller because $\left|E_{\rm B}^{\rm GW}\right|\propto T_{\rm bs}^{-1/4}$, amounting to approximately $6\%$, $14\%$, and $21\%$. 
These estimates indicate that secular cluster evolution can appreciably modify the absolute branching fractions over several relaxation times.
A self-consistent calculation of this evolution would require simultaneously evolving the cluster structure and its BH population, which we do not attempt here.

\section{Gravitational Waves from Binary-Single Scatterings}
\label{sec:results}

In this section, we present numerical Monte Carlo results for the evolution of BH binaries embedded in the NSCs described in \S \ref{sec:cluster_dist}.  One set of simulations uses a fixed cluster concentration $c_\star=6.3$ and varies cluster masses, while the other set fixes cluster mass $M_{\rm tot, \star}=10^{6.3} M_\odot$ and varies concentrations.

\subsection{Monte Carlo Results}

\begin{figure}
\includegraphics[width=80mm]{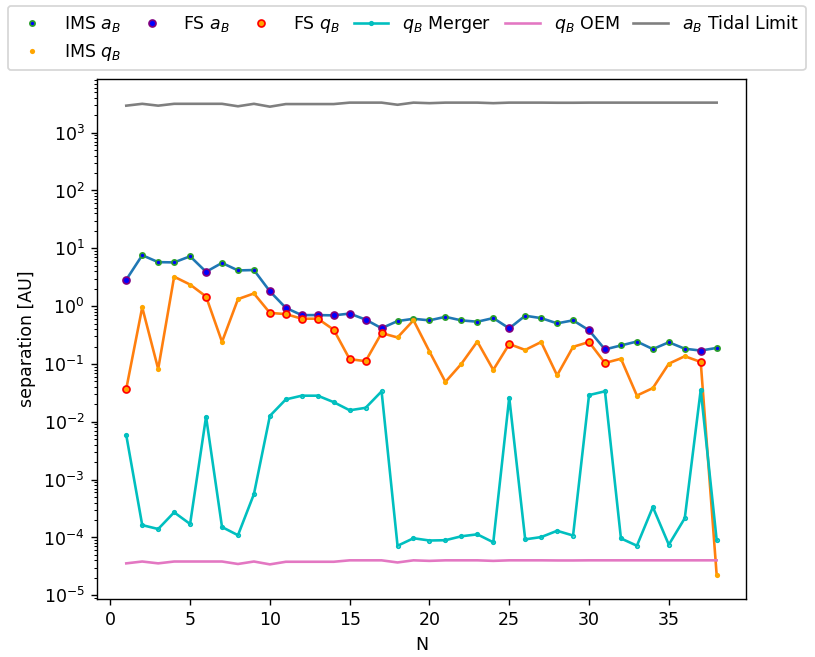}
\caption{Dynamical evolution of an unequal mass BBH in a two-component cluster with \(M_{\rm tot, \star} = 10^{\rm 6.33}, c_{\rm \star} \simeq 6.3\). 
The binary separation, measured in AU, is plotted against the triple state index of the DBS evolution, \(N\) (e.g. $N=20$ represents the binary formed from the 20th metastable triple in the evolutionary sequence). Points with dark blue centers correspond to the binary's semi-major axis $a_{\rm B}$; when circled in green, these are intermediate state $a_{\rm B}$ values, and circled in purple, final state $a_{\rm B}$ values. Points with orange centers correspond to the binary's pericenter \(q_{\rm B}\); when circled in light orange, these are the IMS pericenters, and when circled in red, the FS pericenters.  The light blue line shows the maximum pericenter values needed for a GW-driven merger, and the pink line shows the pericenter values needed for an observably eccentric merger. This particular sequence was selected to illustrate the latter possibility: the final binary in the sequence (which occurs during an IMS) has a pericenter smaller than that needed for an eccentric merger: \(q_{\rm B} < q_{\rm B}^{\rm OEM}\).  This binary therefore merges promptly, with an eccentric imprint in its GW signal.}
\label{fig:MTevolution}
\end{figure}

Performing \(N_{\rm DBS} = 104,000\) dynamical binary sequence evolutions for every cluster model explored, we obtain statistical distributions of the possible outcomes.
In Fig. \ref{fig:MTevolution}, we plot an illustrative example of a single realization of the DBS evolution of a BBH (unequal mass case). This evolutionary sequence contains multiple final states, with intermediate states between them.  It ends in an IMS with a pericenter so small that the binary  undergoes a prompt, observably eccentric merger. This result is exciting, but not a typical outcome: for the cluster parameters illustrated in Fig. \ref{fig:MTevolution}, an observably eccentric merger was only produced in 67 out of the \(104,000\) realizations.  While the outcome of the DBS plotted here is not typical, several representative trends can be observed.  First, we see Heggie's law in action: while the evolution of $a_{\rm B}$ over time is not monotonic, the initially hard binary hardens by roughly one order of magnitude with time.  Second, we see the qualitative similarity between IMS and FS orbital elements: neither binary energy nor angular momentum exhibit systematic differences between IMS and FS binaries.  Finally, we can see a key asymmetry between IMS and FS termination conditions: the maximum pericenter for a GW-driven merger (light blue line) is typically $\approx 2$ orders of magnitude larger in final states than in intermediate states.  This asymmetry stems from timescales: in order for a merger to occur, a FS binary has the long binary-single scattering time to inspiral, while the equivalent IMS binary has only the short tertiary period to inspiral.

\begin{figure}
\includegraphics[width=80mm]{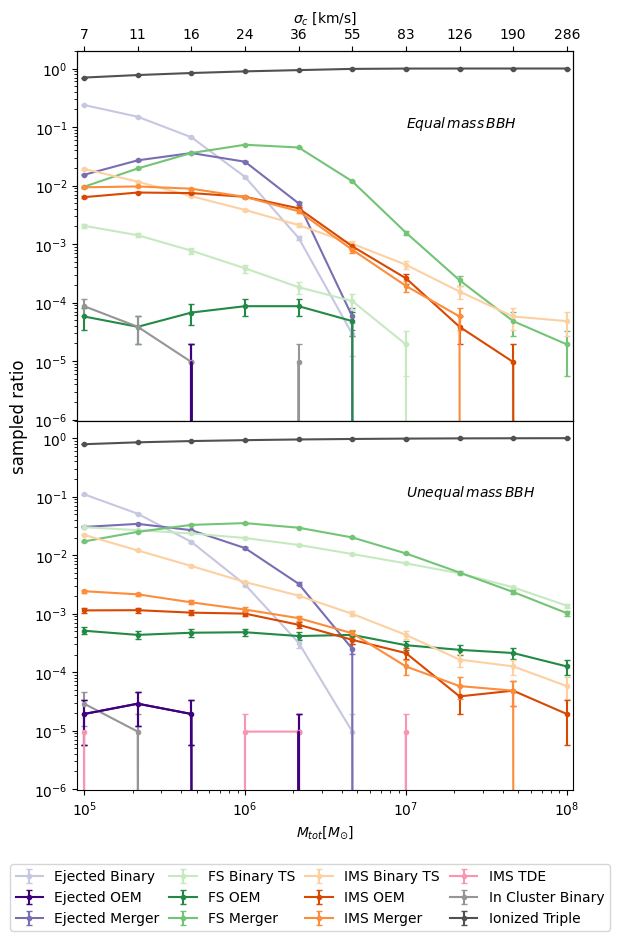}
\caption{Branching ratios of sampled BBH outcomes plotted against varying star cluster total mass, $M_{\rm tot, \star}$, for constant concentration parameter $c_\star = 6.3$. 
The top panel corresponds to the equal mass BBH case, while the bottom panel corresponds to the unequal mass BBH case. For any given cluster mass $M_{\rm tot, \star}$, the sum of all outcome ratios is 1. Different outcomes are color-coded as follows: ionization from three-body scatterings (dark grey), surviving in-cluster binaries (light grey), IMS binary TSs (light orange), FS binary TSs (light green), IMS TDEs (magenta), IMS observably eccentric mergers (dark orange), IMS non-eccentric mergers (orange), FS observably eccentric mergers (dark green), non-eccentric FS mergers (green), ejected non-eccentric mergers (purple), ejected observably eccentric mergers (dark purple), and ejected surviving binaries (light purple). The NSCs we simulate are dynamically hot environments, and so the most common outcome is binary dissolution, usually through direct ionization in a binary single scattering, though occasionally from the closely related pathway of binary TS (which almost exclusively happens to soft binaries gradually growing in semimajor axis).  Final state mergers are always more common than intermediate state mergers, though only by a factor of a few (equal mass BBH case) to $\approx 10$ (unequal mass BBH case) for most of the range.  Conversely, IMSs are more important for the subset of OEM mergers, which may be strongly dominated by IMSs (equal mass BBH case) or see rough parity between IMSs and FSs (unequal mass BBH case). In both panels, there is a critical mass (Eq. \ref{eq:Mejrhcallib}) above which there are no more ejections from the cluster, and the ratio of eccentric mergers to non-eccentric mergers rises.}
\label{fig:sampledratiosMcs}
\end{figure}

\begin{figure}
\includegraphics[width=80mm]{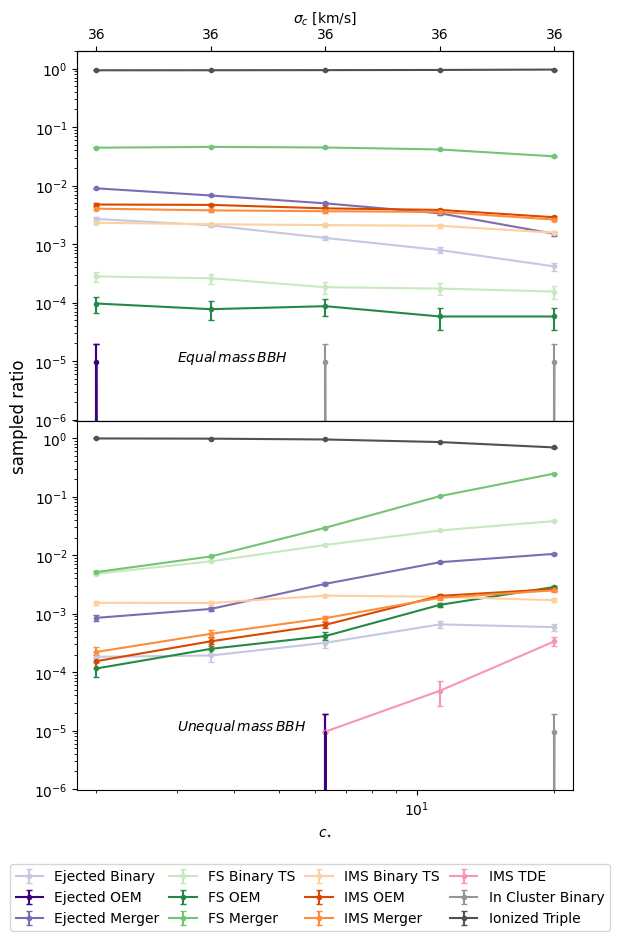}
\caption{Branching ratios of sampled BBH outcomes plotted against varying star cluster concentration parameters \(c_{\rm \star}\), for fixed cluster total mass $M_{\rm tot, \star} = 10^{6.3} M_\odot$. The top panel corresponds to the equal mass BBH case while the bottom panel corresponds to the unequal mass BBH case. The colors of all data points are the same as in Fig. \ref{fig:sampledratiosMcs}. Trends with $c_\star$ are very weak for the equal mass BBH case.  For unequal mass BBHs, we see that most merger-related outcomes increase with increasing $c_\star$.}
\label{fig:sampledratioscs}
\end{figure}

In Fig. \ref{fig:sampledratiosMcs}, we plot branching ratios of different binary evolution outcomes for a range of cluster masses $M_{\rm tot, \star}$, holding cluster concentration $c_\star$ fixed and using the relation in Eq. \ref{eq:rhOfMtot} to set the cluster half-mass radius.  We consider and show results for both the equal mass BBH case and the unequal mass BBH case.  The most common outcome for all binary evolution sequences is ionization, reflecting the broad range of initial semimajor axes we sample (i.e. most sampled binaries are {\it soft} binaries).  A minority of soft binaries break apart due to tidal separation in IMSs or FSs, though the most common breakup mechanism is prompt ionization.

A major trend present in both plot panels is the quick drop-off in the ejected outcomes for \(M_{\rm tot} \gtrsim 10^{6.66} M_\odot\). This can be understood as a cluster mass threshold where the critical semimajor axis for a FS merger, \(a_{\rm GW}\), becomes larger than that for a binary to be ejected from the cluster, \(a_{\rm ej}\). Taking into account the two-component cluster and $M_{\rm tot, \star}-r_{\rm h, \star}$ relation from Eq. \ref{eq:rhOfMtot}, we arrive at the following expression:
\begin{align}
    \label{eq:Mejrhcallib}
    &M_{\rm ej} \simeq M_{\rm l}^{\frac{5}{5+2s}}\left(\frac{m_{\star}}{m_{\rm BH}f_{\rm BH}}\right)^{\frac{10-s}{10+4s}} \left(\frac{5^{\frac{1}{5}}3^{\frac{2}{5}}\pi^{\frac{2}{5}}}{2^{\frac{18}{5}}} \left(\frac{\pi^{2}}{8}-1\right)^{\frac{2}{5}}\right)^{\frac{5s}{5+2s}} \times \\
    & \left(\left(1-\frac{1}{k}\right)\left(\frac{c}{\left(\frac{\rm km}{\rm s}\right)}\right)\frac{m_{\rm a}^{\frac{4}{5}}m_{\rm b}^{\frac{4}{5}}\left\langle m_{\rm s}\right\rangle ^{\frac{4}{5}}}{m_{\rm B}^{\frac{6}{5}}\left(m_{\rm B}+\left\langle m_{\rm s}\right\rangle \right)^{\frac{4}{5}}}\frac{\left(1+c_{\circ}\right)^{\frac{1}{5}}\left(c_{\circ}-1\right)}{\ln\left(c_{\circ}\right)c_{\circ}^{\frac{4}{5}}}\right)^{\frac{5s}{5+2s}}. \nonumber
\end{align}
Here \(k\) is defined as \( \left<E_{\rm B}\right> = k E_{\rm 0}\), and is approximately \(k \simeq 3/2\).
Plugging in the same values that were used for the simulation, we obtain \(M_{\rm ej} \simeq 10^{6.79} M_{\odot}\), in good agreement with the value implied by the MC simulations, \(10^{6.67} - 10^{7.0} M_{\odot}\).
A fuller derivation of \(M_{\rm ej}\) can be found in Appendix \ref{app:Mejder}.

Another notable feature in Fig. \ref{fig:sampledratiosMcs} is the ratio between the IMS OEM rate and the IMS merger rate, which for effectively all \(M_{\rm tot}\) is a number of order unity. We can understand this result analytically by starting from Eqs. \ref{eq:qBEMIMS} and \ref{eq:TGWlessTsExplicit}, rewriting in terms of energy, and using the \(\left<E_{\rm B}\right> = k E_{\rm 0}\) relation again.  We can now approximate the critical pericenter for IMS OEMs to the critical pericenter for all IMS mergers:
\begin{align}
    \label{eq:IMSEMtoIMStotal}
    &\frac{q_{\rm B}^{\rm OEM}}{q_{\rm B}^{\rm merge}} = 0.45\left(\frac{m_{\rm a}}{20M_{\odot}}\right)^{-\frac{1}{7}}\left(\frac{m_{\rm b}}{20M_{\odot}}\right)^{-\frac{1}{7}}\left(\frac{m_{\rm s}}{20M_{\odot}}\right)^{-\frac{1}{7}}\left(\frac{m_{\rm B}}{40M_{\odot}}\right)^{-\frac{5}{21}}  \\
    & \times \left(\frac{f}{10 {\rm Hz}}\right)^{\frac{2}{3}}\left(\frac{E_{0}}{- 364 G M_{\rm \odot}^{2}/{\rm AU}}\right)^{\frac{2}{7}}\left(\frac{k}{0.69}\right)^{-\frac{1}{7}}\left(\frac{1-k}{0.31}\right)^{\frac{3}{7}} \nonumber .
\end{align}
This ratio has been computed by choosing values reasonable for BBH mergers from IMSs, including \(k \simeq 0.69\) and a reference energy scale of \(E_{\rm 0} \simeq -364 G M_{\rm \odot}^{2}/{\rm AU}\).

Another evident trend visible in Fig. \ref{fig:sampledratiosMcs} is the generally steady number of FS OEMs, in contrast to the (larger) number of non-eccentric FS mergers.  The non-eccentric FS merger count declines dramatically for \(M_{\rm tot} > M_{\rm ej}\), while the FS OEM merger count does not. As we consider larger NSC masses, the hard-soft semi-major axis \(a_{\rm HB}\) decreases approximately as \(a_{\rm HB} \propto M_{\rm tot, \circ}^{-2/s}\), increasing the ionized binary fraction and decreasing the number of mergers originating from softer binaries. The decline of the non-eccentric FS merger states therefore indicates these preferentially come from the softer binaries, while the FS OEM population consists of the very hardest binaries, and is thus unaffected by changes in $a_{\rm HB}$.  It is also interesting to note that the FS OEM fraction rises by about half an order of magnitude in the unequal mass BBH case as opposed to the equal mass case, and for the largest NSCs comes to exceed the IMS OEM and IMS merger branching ratios. 

We present an analogous statistical sample of MC results in Fig. \ref{fig:sampledratioscs}, which fixes $M_{\rm tot, \star}=10^{6.3}M_\odot$ and varies the concentration parameter $c_\star$.  In this sample of MC evolutionary sequences, the velocity dispersion is fixed and the central escape velocity changes by very little; as a result, trends with $c_\star$ are generally much weaker than those with $M_{\rm tot, \star}$ that we previously examined in Fig. \ref{fig:sampledratiosMcs}.  For the unequal mass BBH case, we do see a general rise in merger outcome ratios (both eccentric and non-eccentric) as $c_\star$ increases and the binary-single scattering time drops.  

The concentration sequence shown in Fig. \ref{fig:sampledratioscs} also provides a partial test of the sensitivity of our results to secular changes in the cluster environment. Across $c_\star=2$-$20$, the central scatterer number density in our two-component model varies by a factor of approximately $4.6$, while the velocity dispersion remains approximately fixed. Since $T_{\rm bs}\propto\sigma_{\rm c}/n_{\rm c}$, this corresponds to a comparable inverse variation in the encounter timescale and spans a range larger than the factor-of-few change estimated above from balanced expansion over several relaxation times. We emphasize that varying the concentration is not equivalent to evolving a cluster, since secular evolution also changes the velocity dispersion, escape velocity, BH population, and potentially the total cluster mass. Nevertheless, Fig. \ref{fig:sampledratioscs} provides a useful indication of the sensitivity of the branching fractions to changes in the ambient density. In particular, although several individual outcome fractions vary appreciably, especially for unequal mass BBHs, the OEM to total merger fraction is substantially less sensitive to $c_\star$.

\begin{figure*}
\includegraphics[width=160mm]{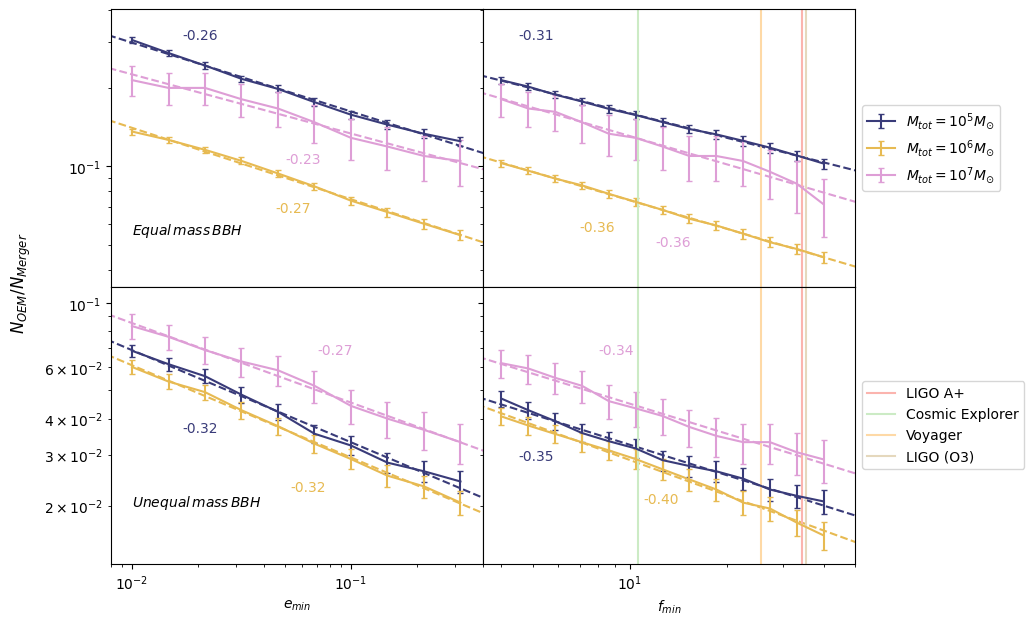}
\caption{Ratios between the number of MC-sampled eccentric mergers ($N_{\rm EOM}$) to the total number of sampled mergers ($N_{\rm merger}$) as a function of eccentricity detection sensitivity. In the left panels, we plot $N_{\rm EOM} / N_{\rm merge}$ against a minimal eccentricity \(e_{\rm f}\) for detection at the fixed frequency \(f = 10\) Hz, while in the right panels, we vary the frequency $f_{\rm min}$ for detection assuming a fixed minimal detectable eccentricity \(e_{\rm min} = 0.1\). The top panels present results of the unequal mass BBH case, and the bottom presents the equal mass BBH case. Different solid curves depict different cluster masses.  
The dashed lines show power-law fits to the MC data, with colored numbers presenting the best-fit power law indices. Vertical lines corresponding to different detectors' expected sensitivities are added to the right panels. We see that the fraction of eccentric mergers to total mergers is in the $\sim$0.01 - 0.10 range, and that the future Cosmic Explorer should roughly double the eccentric merger fraction that LIGO can detect today.}
\label{fig:NEM2MergerMassGrid}
\end{figure*}

Re-evaluating the statistics in our \(M_{\rm tot}\) simulations for their GW implications, we present the ratio of OEMs to total mergers, $N_{\rm EOM}/N_{\rm merge}$, as a function of various eccentricity detectability criteria in Fig. \ref{fig:NEM2MergerMassGrid}. We can see that over all the parameter space of NSC mass \(M_{\rm tot} \in \left[10^5 - 10^7\right] M_{\odot}\) that we explore, this ratio lies within the 0.01-0.10 range for \(f_{\rm min} \in [4, 40] \text{Hz}, e_{\rm f} \in [10^{-2}, 10^{-1/2}]\) (considering the unequal mass BBH case). We also note that for the minimum frequencies detectable by today's LVK collaboration, the percentage of OEMs lies in the 1-4\% range for typical values of NSC \(M_{\rm tot}\). 

We repeat Figs. \ref{fig:sampledratiosMcs} and \ref{fig:sampledratioscs} in Appendix \ref{app:moreMCresults}, but removing all non-merger outcomes.  The resulting Figs. \ref{fig:ratios_Mcs_merger_only} and \ref{fig:ratios_cs_merger_only} more clearly demonstrate the narrow range of variation between numbers of OEMs and total mergers: specifically, the OEM fraction ranges between $\approx 0.01-0.08$.  We also plot the dependence of the OEM fraction on cluster concentration in Fig. \ref{fig:NEM2MergerCsGrid}, finding that the OEM fraction is much less sensitive to $c$ than it is to $M_{\rm tot}$ (Fig. \ref{fig:NEM2MergerMassGrid}).  Finally, we have analyzed the statistics of binaries that underwent an exchange in their DBS. We find no systematic merger outcome is preferred in binaries that undergo exchanges.

\subsection{Comparison to Past Investigations}
\label{subsec:comparison_to_past_investigations}
The OEM fraction\footnote{For simplicity we will use the word "fraction" in this section to indicate the ratio to all mergers. When comparing outcome ratios with other denominators, they will be explicitly stated.} has been studied in the past by several different theoretical methods, for example with a completely analytic approach utilized by \citet{Samsing18}.  In this work, repeated scatterings of equal mass (\(20 M_{\odot}\)) BHs inside a single component, constant BH density environment were used to model the cluster.  The \citet{Samsing18} predictions for the OEM fraction and the ratio of IMS states culminating in in-cluster mergers are qualitatively similar to the results of our MC method, as we show in Table \ref{tab:comparison_with_past_works}.  More generally, this Table provides as close a quantitative comparison as is possible\footnote{Aside from major methodological differences (as we will discuss in this section), past theoretical studies generally made somewhat different assumptions about cluster parameters or BH masses, making the comparisons here useful if somewhat inexact.} between predictions from our work and those from past investigations of prompt GW mergers.

Moving upwards on a scale of numerical complexity, in some other approaches, the cluster environment is modeled by different analytic/semi-analytic distributions (as in this paper), but the non-hierarchical 3-body motion is simulated by N-body codes that integrate the equations of motion (EOM) directly; we will refer to these as \textit{sampled cluster} methods.
This was done for the \(20 M_{\odot}\) equal mass case by \citet{samsing_assembly_2017}: considering incoming singles at the constant speed of \(10 \text{ km/s}\), this work reported a lower bound for the OEM fraction of \(>1\%\). As can be seen in Table \ref{tab:comparison_with_past_works}, the OEM fraction range we find for the comparable clusters in this paper agrees.  
This type of calculation was also done for the unequal mass case by \citet{dallamico_eccentric_2023}, who reported a number of statistics that we compare to our work in Table  \ref{tab:comparison_with_past_works}.  We can see that the OEM fraction and ejected merger fraction of \citet{dallamico_eccentric_2023} are each moderately lower than what we find (by a factor of $\approx 2.5$).  
The ejected OEM fraction and the IMS OEM to total OEM ratios both agree well with the ranges found in this paper.

At a higher, next level of complexity, there are Henon-style Monte Carlo methods \citep{henon_monte_1971}, where whole clusters are evolved simultaneously, using Monte Carlo sampling to advance the Fokker-Plank equation for a spherical, non-rotating star cluster. These codes often directly integrate the EOM during close few-body encounters.  While early works integrated the Newtonian EOM and thus did not account for prompt GW mergers, more recent studies have explored post-Newtonian (PN) dynamics of such few-body scatterings.  For example, \citet{samsing_mocca-survey_2018} re-simulated BH triples extracted from a Monte Carlo cluster simulation using 2.5PN order EOM evolution. The original Monte Carlo simulation explored cluster mass ranges of \(2 \times 10^4 M_{\odot} - 10^5 M_{\odot}\), lower than the NSCs explored here, so we will compare to our lowest mass bin (\(10^5 M_{\odot}\)). As can be seen in Table. \ref{tab:comparison_with_past_works} our OEM fraction and the ratio of IMS mergers occurring inside the cluster to all in-cluster mergers both agree well with \citet{samsing_mocca-survey_2018}. 

Other cluster Monte Carlo approaches have incorporated PN EOM evolution directly into the code \citep{rodriguez_post-newtonian_2018_formation, rodriguez_post-newtonian_2018_highly_eccentric}.  For example,  \citet{rodriguez_post-newtonian_2018_highly_eccentric} simulate globular clusters with masses \(0.4\times 10^5 M_{\odot} - 6.7 \times 10^5 M_{\odot}\), comparable to our \(M_{tot} \in [10^5 - 10^6] M_{\odot}\) range. They report an OEM fraction, an ejected merger fraction, an IMS in in-cluster merger ratio and a total IMS merger fraction that all agree well with the statistics found in this paper.
\citet{rodriguez_post-newtonian_2018_formation} explore a similar range of cluster masses, between \(10^5 - 1.2\times 10^6\), again comparable to our \(10^5 M_{\odot} - 10^{6.33} M_{\odot}\) bins. As can be seen in Table \ref{tab:comparison_with_past_works}, they report a slightly higher OEM fraction than the one found in this paper for both the \(e_{\rm min} = 0.1\), \(e_{\rm min} = 0.05\) detection sensitivity regimes. They also report a range for the IMS merger fraction that overlaps with the range found in this paper.

Finally, and most realistically, there are direct N-body simulations, which evolve the EOM of the whole cluster, although prescriptions and sub-grid models are sometimes used for various non-dynamical phenomena such as stellar evolution, CO formation, and common envelope phases. A limitation of this approach is the high computational expense, and as a result the cluster masses explored are usually not very high.  For example, \citet{di_carlo_merging_2019} simulates clusters in the mass ranges of \(10^3 - 3\times 10^4 M_{\odot}\), consistent with young open clusters. They report an OEM fraction of only \(\sim 2.5 \%\) when considering the minimum detectable eccentricity as \(e_{\rm min} = 10^{-2}\) at orbital frequency of \(f_{\rm min} = 10^{-2}\), much smaller than the \((38.7 \pm 0.7)\%\) that we find for our smallest mass clusters of \(10^5 M_{\odot}\) at similar detection sensitivity. 
This may be due to the absence of post-Newtonian terms in the \citet{di_carlo_merging_2019} equations of motion, which will make it impossible for intermediate state binaries to undergo prompt GW-driven inspirals. Considering most mergers in the \citet{di_carlo_merging_2019} are ejected, we compare their reported OEM fraction to the ratio of our ejected OEMs to total ejected mergers. We find order of magnitude difference that might be reconciled after adjusting the cluster masses.

\begin{table*}
    \centering
    \begin{tabular}{|m{7em}|m{10em}|m{8em}|m{9em}|m{6em}|m{12em}|}
        \hline
        Comparison paper (CP) & Cluster in CP & Closest clusters in this paper & Result Name & Values in CP & Values in this paper \\
        \hline
        \hline
        \citealt{samsing_assembly_2017} (Sampled cluster) & Equal mass \((20 M_{\odot})\) \(v_s = 10 \text{km/s}\) &Equal mass, \(\sigma_{c} \in  [7 - 11] \text{ km/s} \) & OEM fraction & \(>1\%\) & \((12.0 \pm 0.4)\% - (15.7 \pm 0.6)\%\) \\
        \hline
        \multirow{3}{7em}{\citealt{samsing_mocca-survey_2018} (He’non-style Monte Carlo)}  & \multirow{3}{10em}{Unequal mass, \(M_{\rm tot} \in  [2\times 10^4, 10^5] M_{\odot}\)} & \multirow{3}{8em}{Unequal mass, \(M_{tot} = 10^5 M_{\odot}\)} & \multirow{2}{9em}{OEM fraction} & \multirow{2}{6em}{\(1\% - 5\%\)} &\multirow{2}{12em}{\((3.2 \pm 0.2)\%\)} \\
        & & & & & \\
        \cline{4-6}
        & & & IMS in in-cluster mergers & \(\sim 10 \%\) & \((16.7 \pm 0.9)\%\) \\
        \hline
        \multirow{4}{7em}{\citealt{rodriguez_post-newtonian_2018_highly_eccentric} (He’non-style Monte Carlo)} & \multirow{4}{10em}{Unequal mass, \(M_{\rm tot} \in [0.4 - 6.7] \times 10^5 M_{\odot}\)} & \multirow{4}{8em}{Unequal mass, \(M_{tot} \in [10^5 - 10^6] M_{\odot}\)} & OEM fraction & \(3\%\) & \((2.4 \pm 0.2)\% - (3.2 \pm 0.2)\%\) \\
        \cline{4-6}
        & & & Ejected merger fraction & \((45\% - 55\%)\) & \((25.8 \pm 0.8)\% - (59 \pm 1)\%\) \\
        \cline{4-6}
        & & & IMS in in-cluster mergers & \(10\%\) & \((5.7 \pm 0.4)\% - (16.7 \pm 0.9)\%\) \\
        \cline{4-6}
        & & & IMS merger fraction &  \(3 \%\) &  \((4.1 \pm 0.3)\% - (6.8 \pm 0.4)\%\) \\
        \hline
        \multirow{2}{7em}{\citealt{Samsing18} (Analytic)}  & \multirow{2}{10em}{Equal mass \((20 M_{\odot})\)}& \multirow{2}{8em}{Equal mass, \(M_{\rm tot} \in  [10^5 - 10^7] M_{\odot}\)} & OEM fraction & \(5\%\) & \((7.1 \pm 0.7)\% - (15.8 \pm 0.6)\%\) \\
        \cline{4-6}
        &  &  & IMS in in-cluster mergers & \(10\%\) & \((13 \pm 1)\% - (62 \pm 2)\%\)  \\

        \hline
        \multirow{3}{7em}{\citealt{rodriguez_post-newtonian_2018_formation} (He’non-style Monte Carlo)} & \multirow{3}{10em}{Unequal mass, \(M_{\rm tot} \in  [10^5, 1.2\times 10^6] M_{\odot}\)} & \multirow{3}{8em}{Unequal mass, \(M_{tot} \in [10^5 - 10^{6.33}] M_{\odot}\)} & OEM fraction & \(4\%\) & \((2.4 \pm 0.2)\% - (3.3 \pm 0.2)\%\) \\
        \cline{4-6}
        & & & OEM fraction with \(e_{\rm min} = 0.05\) & \(6\%\) & \((3.0 \pm 0.2)\% - (4.1 \pm 0.3)\%\) \\
        \cline{4-6}
        & & & IMS merger fraction & \(0.25\% - 5\%\) & \((4.1 \pm 0.3)\% - (6.8 \pm 0.4)\%\) \\
        \hline
        
        \multirow{3}{7em}
        {\citealt{di_carlo_merging_2019} (Direct N-body)} & \multirow{3}{10em}{Unequal mass, \(M_{\rm tot} \in  [10^3, 3\times 10^4] M_{\odot}\)}& \multirow{3}{8em}{Unequal mass, \(M_{tot} = 10^5 M_{\odot}\)} & OEM fraction of \(e_{\rm min}=10^{-2}, f_{\rm min}=10^{-2}\) & \(\sim 2.5 \%\) & \((38.7 \pm 0.7)\%\) \\
        \cline{4-6}
        & & & Ejected merger fraction & \(\sim 98\%\) & \((59 \pm 1)\%\) \\
        \cline{4-6}
        & & & Ejected OEMs in ejected mergers with \(e_{\rm min}=10^{-2}, f_{\rm min}=10^{-2}\) & \(\sim 2.5 \%\) & \((11.7 \pm 0.6)\%\) \\
        
        \hline
                
        \multirow{4}{7em}{\citealt{dallamico_eccentric_2023} (Sampled cluster)} & \multirow{4}{10em}{Unequal mass, \(\sigma_{c} = 20 \text{ km/s}\)} & \multirow{4}{8em}{Unequal mass, \(\sigma_{c} \in  [16 - 24] \text{ km/s} \)} & OEM fraction & \(1 \%\) & \((2.4 \pm 0.2)\% - (2.9 \pm 0.2)\%\) \\
        \cline{4-6}
        & & & Ejected merger fraction & \(18 \%\) & \((26.8 \pm 0.7)\% - (42.6 \pm 0.9)\%\) \\
        \cline{4-6}
        & & & Ejected OEM fraction & \(0\%\) & \(0\% - (0.03 \pm 0.02)\%\) \\
        \cline{4-6}
        & & & IMS OEM to total OEM &  \(67 \%\) & \((67 \pm 8)\% - (68 \pm 8)\%\) \\
        \hline
        
    \end{tabular}
    \caption{Summary of comparison to past works (\S \ref{subsec:comparison_to_past_investigations}). Each paper this work was compared to explored different cluster parameters (2nd column) and in turn was compared to different cluster parameters explored in this paper (3rd column). The result names are as follows: If not indicated otherwise, OEMs are calculated with \(e_{\rm min} = 0.1, f=10 \text{Hz}\). The word fraction is used to indicate a ratio to all mergers, so {\it OEM fraction}, {\it IMS Merger fraction}, {\it Ejected Merger fraction} and {\it Ejected OEM fraction} indicate the ratio of the OEM, IMS, ejected and ejected OEM mergers to all mergers, respectively. {\it IMS OEM to total OEM} indicates the ratio of OEMs in IMSs to all OEMs.  {\it IMS in in-cluster mergers} indicates the ratio of IMS mergers inside the cluster to all mergers happening inside the cluster (i.e. not ejected). {\it Ejected OEMs in ejected mergers} indicates the ratio between the ejected OEMs to all ejected mergers.
    We find an order of magnitude agreement when comparing the OEM fraction to other works examining similar cluster parameters. Uncertainties are calculated by the approximated standard deviation (\(\sqrt{p\left(1-p\right)/N}\) where \(p\) is the sampled probability) of outcomes measured in DBS simulations, with standard uncertainty propagation. }
    \label{tab:comparison_with_past_works}
\end{table*}

We have introduced a new method for modeling transient, non-hierarchical three-body systems formed within stellar clusters. Similar to the "sampled cluster" methods from previous literature that we described in this section, our method models clusters via analytic distributions. 
The novelty of our approach is that the three-body motion is also treated with analytic distributions, providing more direct physical insight, opportunities to derive relevant scaling relations, and of course a much reduced computational expense compared to more numerical methods. Another advantage of this approach, not exploited in this paper, is its ability to incorporate highly non-spherical geometries: for example, repeated binary-single scatterings in geometrically thin disks of stars may play an important role in creating GWS in active galactic nuclei \citep{Stone+17, Gilbaum+25}, with the quasi-planar geometry potentially leading to very super-thermal binary eccentricity distributions \citep{samsing_agn_2022}.  These highly non-spherical environments cannot be simulated by existing Henon-style cluster Monte Carlo codes, though they are accessible to direct N-body simulations.  The main disadvantages of our method are its reliance on strong assumptions about environmental geometry and stellar distributions, and its lack of explicit time evolution.  Another disadvantage is that our specific three-body modeling presently limits us in the physics that can arise during binary single scatterings: we only incorporate ergodic outcome distributions for strong non-hierarchical systems.  While weaker scatterings, either impulsive \citep{ModakHamilton23} or secular \citep{HamersSamsing19, leigh_thermodynamics_2022}, could eventually be incorporated into our modeling, the quantitative addition of strong but non-chaotic (e.g. prompt exchange outcomes) would require further theoretical development.

\section{Conclusions}
\label{sec:conclusions}
In this paper, we have investigated the elliptic distribution of the chaotic three-body problem's outcome states, which were first derived in \citet{GinatPeretz2021}, and which characterize intermediate states of quasi-regular evolution during three-body chaos \citep{Samsing14}.  We found several novel features of elliptic outcome distributions, and used these (along with the analogous hyperbolic outcome distribution \citealt{StoneLeigh19}) to describe the evolution of a binary system undergoing scatterings with an incoming single star. We then developed a Monte Carlo scheme to sample the initial conditions of binaries and incoming singles assuming a two-component cluster model (with a black hole subsystem). We used this algorithm, which draws elliptic and hyperbolic outcomes from analytic distributions using the same Monte Carlo sampling technique, to evolve binaries within clusters of different parameters.  We explored a broad cluster parameter space along the empirical \(M_{\rm tot, \star} - r_{\rm h, \star}\) relation (Eq. \ref{eq:rhOfMtot}). Our results are summarized as follows:

\begin{enumerate}
    \item {\it We introduced a dimensionless parameter $\beta$ to distinguish between chaotic scrambles and temporarily hierarchical excursions (i.e. periods of quasi-regular motion).  We found that the best agreement with numerical scattering experiments is achieved when $\beta=\alpha$, the free parameter previously used to define the chaotic region in outcome space.} This differs from the previous work by \citet{GinatPeretz2021}, where the two parameters used for the model are independent. The elliptic three-body outcome distribution is re-derived in Eq. \ref{eq:3ddist} and is given as a function of binary orbital elements as opposed to the form of \citet{GinatPeretz2021}, which is given in angular momentum components. When incorporating different chaotic region boundaries and comparing scramble number probabilities to numerical results, we find the numerical values fit \(\beta=\alpha=2.5\) and \(\beta=\alpha=3.5\) for simple and apocentric escape criteria, respectively. Explicitly demanding the tertiary's pericenter to be inside the chaotic region rids the phase space volumes of their imaginary parts, and the analytic boundaries for the simple escape criteria conserving this demand are derived (see App. \ref{app:phasespacebounds}).
    \item {\it We explored the marginalizations of the elliptic and hyperbolic outcome distributions under different chaotic region criteria and find they are tightly linked.} The elliptic and hyperbolic \({\rm d}\sigma/{\rm d}E_{\rm B}\) distributions share a common boundary when \(E_{\rm B} \rightarrow E_{\rm 0}\), which is given analytically in Eq. \ref{eq:E0limit} and shown graphically in Fig. \ref{fig:Energy1Ddist}. The intermediate state eccentricity distribution \({\rm d}\sigma/{\rm d}e_{\rm B}\) is neither thermal nor super-thermal for mid-range \(e_{\rm 0}\) values. A second-order discontinuity in the \({\rm d}\sigma/{\rm d}e_{\rm B}\) distribution is identified and its analytic origin in the \((E_{\rm B}, e_{\rm B}, C_{\rm B} = 1)\) phase plane is obtained in Eq. \ref{eq:Spike}. The cosine of the inclination distribution \({\rm d}\sigma/{\rm d} C_{\rm B}\) is universal up to \(10\%\) deviation when considering elliptic, hyperbolic, simple escape or apocentric escape chaotic region criteria and different angular momenta. An analytic form for the bivariate intermediate state outcome distribution, \({\rm d}\sigma_{\rm B}/{\rm d}E_{\rm B}{\rm d}L_{\rm B}\), integrated over \(C_{\rm B}\), is given in Appendix \ref{app:CB}.
    \item {\it We integrated over both distributions to obtain disintegration, escape and excursion probabilities.  We compare the probability the system will undergo N scrambles with numerical simulations; we find good agreement as N rises in high \(L_{\rm 0}\) systems.} For \(m_{\rm 1} = m_{\rm 2}\) systems, the probability for an \(m_{\rm 3} < m_{\rm 1}\) particle to escape (Eq. \ref{eq:Pm3es}), given an escape of any mass, is larger than the probability for it to have an excursion given any excursion (Eq. \ref{eq:Pm3ex}).  The opposite is true for \(m_{\rm 3} > m_{\rm 1}\). The behavior of escape and excursion probabilities is qualitatively similar for any \(e_{\rm 0}\) values. For the same \(m_{\rm 1} = m_{\rm 2}\) systems, the disintegration probability (Eq. \ref{eq:Pdis}) is lowest when \(m_{\rm 3} = m_{\rm 1} = m_{\rm 2}\). It rapidly grows to 1 as \(m_{\rm 3}/m_{\rm 1} \rightarrow 0 \) and goes to a constant for \(m_{\rm 3}/m_{\rm 1} \gg 1\).
    \item {\it We designed a Monte Carlo algorithm to model repeated scatterings in a two-component star cluster, and applied this model to run large statistical samples of iterated binary-single scatterings in nuclear star cluster type environments}. We classified different binary black hole evolution outcomes, across cluster mass ranges of \( M_{\rm tot} \in [10^5 - 10^8] M_{\odot} \). We find that there exists a critical total cluster mass \(M_{\rm ej}\) in the range of \(10^{6.67} - 10^{7.0} M_{\odot}\) above which black hole binaries that would otherwise be destined for ejection merge first, in the cluster, and no ejected binaries escape. An analytic calculation for this critical mass is presented in Appendix \ref{app:Mejder}. Given our focus on modeling intermediate states, we pay special attention to the production of intermediate states and observably eccentric binary black hole mergers, using approximate cutoffs for minimal detectable eccentricity ($e_{\rm f}=0.1$ at $f=10~{\rm Hz}$).  We find the ratio of intermediate state observably eccentric mergers to all intermediate state mergers is about \((40 \pm 2)\% - (57 \pm 13)\%\) for equal masses and \((32 \pm 3)\% - (62 \pm 17)\%\) for unequal masses, and propose a scaling relation to approximate this fraction in Eq. \ref{eq:IMSEMtoIMStotal}. Next, we find that the fraction of observably eccentric mergers in final states (binaries left behind after triple disolution) is comparable to the intermediate state observably eccentric merger fraction in the unequal mass case, in contrast to a 1-2 order of magnitude difference in the equal mass case, which may lead completely analytic approaches to underestimate its significance. For \(M_{\rm tot} > M_{\rm ej}\) there is a decline in the number of final state mergers which in turn corresponds to an increasing observably eccentric merger fraction. We connect this decline with the decrease of the hard-soft boundary. We vary the cluster compactness parameter in the range of \(c_{\star} \in [2 - 20]\) for a typical \(M_{\rm tot} = 10^{6.33} M_{\odot}\) cluster and find no statistically significant impact that persists in both the equal mass and unequal mass cases.
    \item {\it  We found that the observably eccentric merger to total merger fraction for our fiducial eccentricity detection criterion (\(f_{\rm min}=10\text{ Hz}, e_{\rm min} = 0.1\) ) is in the \((2.4 - 4.4)\%\) range for unequal mass binary black holes, across the full landscape of cluster masses \(M_{\rm tot}\) that we explored }. We then relaxed our assumptions of minimum detectable eccentricity and frequency and re-evaluated the observable eccentric merger fraction for all simulations. For any threshold frequency \(f_{\rm min} \in [4, 40] \text{ Hz}\) and any threshold \(e_f \in [10^{-2}, 10^{-1/2}]\), the observably eccentric merger fraction lies in the broader \((1.5 - 8.3)\%\) range. The observably eccentric merger fraction in the unequal mass range changes to \((1.6 - 3.1)\%\), \((1.8 - 3.3)\%\), \((2.4 - 4.3)\%\) when taking into account the frequency sensitivities of LIGO \((f_{\rm min} \simeq 34.4 \text{ Hz})\), Voyager \((f_{\rm min} \simeq 25.5 \text{ Hz})\) and Cosmic Explorer \((f_{\rm min} \simeq 10.6 \text{ Hz})\), respectively. This prediction is compatible with the fraction of eccentric mergers observed by LIGO \((0 - 4.4)\%\) until now. 
    Order of magnitude or better agreement is found when comparing the observably eccentric merger fraction to other works examining similar cluster masses (see Table \ref{tab:comparison_with_past_works}).
\end{enumerate}

While our modeling of intermediate state microcanonical ensemble has revealed novel features of the intermediate state outcome distributions, and has shown its potential for astrophysical applications (e.g. estimating eccentric binary black hole merger rates), it has a number of limitations, the most important of which we briefly recapitulate here.  
Although the \(\alpha\) and \(\beta\) factors come from intuitive demands on the chaotic region, they remain approximations rather than first principle evaluations of the problem. 
Another limitation of our approach is that no information can be extracted about the scrambles, and thus our sequence of binary states is ultimately a discrete-time approximation to a continuous-time reality.  
To incorporate the outcome distributions into our Monte Carlo algorithm, we assume that the three-body phase space has already been ergodicized since the first scramble; equivalently, interactions that are not completely chaotic are not taken into account. 
Long-lived three-body systems interacting with an incoming single along with binary-binary interactions within clusters necessitate the need for a similar handling of 4-body interactions which was out of the scope of this paper. 
The Monte Carlo evolution of a binary black hole after merger is neglected in our binary evolution sequences; while it is possible that this is a good approximation to reality (because of gravitational wave recoil kicks), it neglects the possibility of merger product retention and hierarchical growth.  
Finally, our Monte Carlo scheme's treatment of the background cluster is flexible but for now quite approximate.

Improving upon all these limitations may be subjects of future works, but exploring the additional gravitational wave merger recoil kick due to individual mass spins and its effects on merger product retention would be the most interesting. If one finds a preferential cluster parameter space for repeated black hole mergers, and consequently clusters that can grow black holes into the pair-instability mass gap and beyond, this would immediately inform not just LVK data analysis but also open questions about intermediate-mass black hole formation. Recently, such an intermediate-mass black hole of mass \(8,200 M_{\odot}\) was found in Omega Centauri, a large cluster of total mass \(\sim 10^{6.6} M_{\odot}\)
\citep{haberle_fast-moving_2024}. 
Another interesting future direction is the adaptation of our model to 3-body interactions within discs. It has been proposed that dynamical encounters in discs around supermassive black holes may drive up the observably eccentric merger fraction in stellar mass binary black holes (\citealt{trani_three-body_2024}, \citealt{samsing_agn_2022}). While the usual Henon style codes assume spherical symmetry and cannot be adapted for such exploration, the framework presented in this paper might be flexible enough to explore this parameter space. 
For example, a completely analytic approach using the elliptic and hyperbolic distributions has already been employed to characterize the hard-soft boundary in different environments including discs \citep{ginat_perets_soft_binaries_2021}.

Overall, the treatment of chaotic three-body motion using a statistical microcanonical ensemble \citep{monaghan_statistical_1976, monaghan_statistical_1976-1, valtonen_three-body_2006, StoneLeigh19, ginat_binaries_2021, ginat_analytic_2022} has proven to be a useful tool in probing the nature of the three-body problem as well as the origins of gravitational wave mergers. We hope that it may also be further used and improved to better understand the masses of the black hole population in different stellar environments, and via comparison to other notions of ergodicity \citep{kol_flux-based_2021}, to better refine our understanding of chaos in self-gravitating systems.

\section*{Acknowledgements}
DM and NCS were supported by the Israel Science Foundation (Individual Research Grant 2565/19) and the Binational Science Foundation (grant Nos. 2019772 and 2020397).  
NWCL gratefully acknowledges the generous support of a Fondecyt General grant 1230082, as well as support from Millenium Nucleus NCN2023\_002 (TITANs) and funding via the BASAL Centro de Excelencia en Astrofisica y Tecnologias Afines (CATA) grant PFB-06/2007.  NWCL also thanks support from ANID BASAL project ACE210002 and ANID BASAL projects ACE210002 and FB210003.

\section*{Data Availability}
\label{sec:code_data_avail}
We hereby introduce the open source code \texttt{probab3} used to generate the dynamical binary sequence (DBS) evolution statistics as described in this paper. 
\texttt{probab3} code is available as a Python package, a command-line tool and in raw repository form at \texttt{https://github.com/DinaMeylakh/probab3}.

Although it can be made faster by using a compiled language or utilizing GPU for the integration steps, we opted to use Python for traceability, readability, and its fast-growing community. 

Other methods to make the code faster exist, like replacing the pre-evolution with a deterministic algorithm for initial conditions or pre-tabulating the integration outcomes of Eq. \ref{eq:3ddist} and its hyperbolic analog. Nonetheless, we found an average of 17 s per equal mass DBS and 55 s per unequal mass DBS satisfactory when running multiple samples simultaneously.

The data underlying this paper can be re-generated using \texttt{probab3} and will be shared on reasonable request to the corresponding author.

\bibliographystyle{mnras}
\bibliography{IMS_refs}

\appendix

\onecolumn

\section{Simplifications and Limiting Cases of \(\sigma\)} 
\label{app:CB}

If we rewrite Eq. \ref{eq:3ddist} in terms of binary angular momentum \(L_{\rm B}\), binary energy \(E_{\rm B}\) and the cosine of the inclination \(C_{\rm B}\) in the following manner:
\begin{align}
&\sigma = \frac{2\pi^4G^2M^{\frac{5}{2}}m_{\rm B}}{\left(m_{\rm a}m_{\rm b}m_{\rm s}\right)^{\frac{3}{2}}}\int\int\int\frac{L_{\rm B}{\rm d}E_{\rm B}{\rm d}L_{\rm B}{\rm d}C_{\rm B}}{L_{\rm s}\left(E_{\rm B}-E_{\rm 0}\right)^{\frac{3}{2}}\left(-E_{\rm B}\right)^{\frac{3}{2}}} \nonumber \\
&\times\left(\cos^{-1}\left(\frac{1-\frac{2R}{Gm_{\rm B}m_{\rm s}}\left(E_{\rm B}-E_{\rm 0}\right)}{\sqrt{1-\frac{2M\left(E_{\rm B}-E_{\rm 0}\right)L_{\rm s}^2}{G^2m_{\rm B}^3m_{\rm s}^3}}}\right)-\sqrt{\frac{2M\left(E_{\rm B}-E_{\rm 0}\right)}{G^2m_{\rm s}^3m_{\rm B}^3}\left(2RGMm^2-2m\left(E_{\rm B}-E_{\rm 0}\right)R^2-L_{\rm s}^2\right)} \right),
\label{eq:3ddistAngularMomentum}
\end{align}
then we can further analytically integrate the 3d distribution over \(C_{\rm B}\).
Denoting the integral over \(C_{\rm B}\) as \(g\):
\begin{align}
&g = \frac{2\pi^4G^2M^{\frac{5}{2}}m_{\rm B}}{\left(m_{\rm a}m_{\rm b}m_{\rm s}\right)^{\frac{3}{2}}}\int\frac{L_{\rm B}{\rm d}C_{\rm B}}{\left(L_{\rm 0}^2-2L_{\rm 0}L_{\rm B}C_{\rm B} +L_{\rm B}^2\right)^{\frac{1}{2}}\left(E_{\rm B}-E_{\rm 0}\right)^{\frac{3}{2}}\left(-E_{\rm B}\right)^{\frac{3}{2}}} \nonumber \\
&\times\left(\cos^{-1}\left(\frac{1-\frac{2R}{Gm_{\rm B}m_{\rm s}}\left(E_{\rm B}-E_{\rm 0}\right)}{\sqrt{1-\frac{2M\left(E_{\rm B}-E_{\rm 0}\right)\left(L_{\rm 0}^2-2L_{\rm 0}L_{\rm B}C_{\rm B} +L_{\rm B}^2\right)}{G^2m_{\rm B}^3m_{\rm s}^3}}}\right)-\sqrt{\frac{2M\left(E_{\rm B}-E_{\rm 0}\right)}{G^2m_{\rm s}^3m_{\rm B}^3}\left(2RGMm^2-2m\left(E_{\rm B}-E_{\rm 0}\right)R^2-\left(L_{\rm 0}^2-2L_{\rm 0}L_{\rm B}C_{\rm B} +L_{\rm B}^2\right)\right)} \right)
\label{eq:3ddistAngularMomentumIntegral}
\end{align}
evaluating the integral, we find
\begin{align}
    &g = \frac{\pi^4G^2m_{\rm B}M^{\frac{5}{2}}}{\left(-E_{\rm B}\right)^{\frac{3}{2}}L_{\rm 0}\left(E_{\rm B}-E_{\rm 0}\right)^{\frac{3}{2}}{\left(m_{\rm a}m_{\rm b}m_{\rm s}\right)^{\frac{3}{2}}}}\left(2L_{\rm s}\sqrt{\frac{\left(E_{\rm B}-E_{\rm 0}\right)R^2\Delta E_{\rm s}}{G}}
    -2L_{\rm s}\cos^{-1}\left(\frac{\frac{2R\left(E_{\rm B}-E_{\rm 0}\right)}{Gm_{\rm s}m_{\rm B}}+1}{\sqrt{\frac{2\left(E_{\rm B}-E_{\rm 0}\right)L_{\rm s}^2M}{G^2m_{\rm s}^3m_{\rm B}^3}+1}}\right)\right.\nonumber \\
    & \left. -\frac{Gm^{\frac{3}{2}}M\sqrt{2{\rm sign}\left(E_{\rm B}-E_{\rm 0}\right){\rm sign}\left(\Delta E_{\rm s}\right)}}{{\rm sign}\left(E_{\rm B}-E_{\rm 0}\right)\sqrt{\left|E_{\rm B}-E_{\rm 0}\right|{\rm sign}\left(\Delta E_{\rm s}\right)}}\tanh^{-1}\left(\frac{L_{\rm s}\left(GmM+2R\left(E_{\rm B}-E_{\rm 0}\right)\right)}{G m M\sqrt{2R^2m\Delta E_{\rm s}}}\right) \right. \nonumber \\
    & \left. +\frac{\sqrt{2}\left(G^2m^2M^2-2R^2\left(E_{\rm B}-E_{\rm 0}\right)^2\right)}{GMm\sqrt{\left(E_{\rm B}-E_{\rm 0}\right)}}\log\left(\left(E_{\rm B}-E_{\rm 0}\right)M\left(L_{\rm s}+{\rm sign}\left(E_{\rm B}-E_{\rm 0}\right)\sqrt{2R^2m \Delta E_{\rm s}}\right)\right)\right).
    \label{eq:2ddistAngularMomentum}
\end{align}
Here \(\Delta E_{\rm s} = E_{\rm crit} - \left(E_{\rm 0} -E_{\rm B}\right)\). So the integrated phase space volume becomes
\begin{align}
    \sigma = \int \int \left[ g\left(E_{\rm B},L_{\rm B },C_{\rm B} \right)\right]_{C_{\rm B}^{-}}^{C_{\rm B}^{+}}{\rm d}E_{\rm B}{\rm d}L_{\rm B}
\end{align}
when integrating over \(C_{\rm B} \in [C_{\rm B}^{-}, C_{\rm B}^{+}]\).  While we were not able to identify a further simplification to produce 1d marginal distributions, this 2d bivariate distribution has not, to the best of our knowledge, appeared previously in the literature.  While it is somewhat more unwieldy than the original 3d (trivariate) intermediate state outcome distribution that it stems from, it may nonetheless prove useful in e.g. studies of intermediate states in which orientation is a nuisance parameter that must be integrated over.

\section{Analytic phase space bounds}
\label{app:phasespacebounds}
We start off with the requirement that the teritiary's pericenter lie inside the chaotic region, as in Eq. \ref{eq:qslessR}. Re-writing this requirement using energies and angular momenta we arrive at Eq. \ref{eq:StartForBounds}. Solving this inequality for \(L_{\rm B}\) we obtain the integration bounds for which \(L_{\rm B} \in \left[L_{\rm B, -}, L_{\rm B, +}\right]\):
\begin{align}
    L_{\rm B, \pm} = L_{\rm 0}C_{\rm B}\pm\sqrt{L_{\rm 0}^{2}\left(C_{\rm B}^{2}-1\right)+2RGMm^{2}+2\left(E_{\rm 0}-E_{\rm B}\right)R^{2}m}.
    \label{eq:LBverge}
\end{align}
We can see that the \(L_{\rm B}\) domain is non-zero only when the discriminant in \ref{eq:LBverge} is positive, i.e. 
\begin{align}
    L_{\rm 0}^{2}\left(C_{\rm B}^{2}-1\right)+2RGMm^{2}+2\left(E_{\rm 0}-E_{\rm B}\right)R^{2}m > 0.
    \label{eq:LBvergeDetgeq0}
\end{align}
Using the simple escape criteria from Eq. \ref{eq:simpleesc} and rearranging we get an equivalent demand on \(E_{\rm B}\):
\begin{align}
    0 < \frac{2L_{\rm 0}^{2} \left(C_{\rm B}^{2}-1\right)E_{\rm B}^{2}}{m\alpha^{2}G^{2}m_{\rm a}^{2}m_{\rm b}^{2}}-\left(\frac{2Mm}{\alpha m_{\rm a}m_{\rm b}}+1\right)E_{\rm B}+E_{\rm 0}
    \label{eq:LBvergeDetgeq0EB}
\end{align}
Solving this inequality for \(E_{\rm B}\), we obtain the integration bounds for which \(E_{\rm B} \in \left[E_{\rm B, +}, E_{\rm B, -}\right]\):
\begin{align}
    E_{B,\pm} = \frac{m\alpha Gm_{\rm a}m_{\rm b}}{4L_{\rm 0}^{2}\left(C_{\rm B}^{2}-1\right)}\left(G\left(2Mm+\alpha m_{\rm a}m_{\rm b}\right)\pm\sqrt{G^{2}\left(2Mm+\alpha m_{\rm a}m_{\rm b}\right)^{2}-8\frac{L_{\rm 0}^{2}\left(C_{\rm B}^{2}-1\right)E_{\rm 0}}{m}}\right).
    \label{eq:EBverge}
\end{align}
Again, we see that the \(E_{\rm B}\) domain is non-zero only when the discriminant of \ref{eq:EBverge} is positive. This translates to the following inequality:
\begin{align}
    C_{\rm B}^{2} > 1+\frac{G^{2}\left(2Mm+\alpha m_{\rm a}m_{\rm b}\right)^{2}m}{8L_{\rm 0}^{2}E_{\rm 0}}
    \label{eq:EBvergeDetgeq0CB}
\end{align}
Solving this inequality for \(C_{\rm B}\), we obtain the integration bounds for which \(C_{\rm B} \in \left[-1, C_{\rm B, -}\right] \bigcap \left[C_{\rm B, +}, 1\right]\):
\begin{align}
    C_{\rm B,\pm} = \pm\sqrt{ 1 + \frac{G^{2}\left(2Mm+\alpha m_{\rm a} m_{\rm b} \right)^{2}m}{8L_{\rm 0}^{2}E_{\rm 0}}}.
    \label{eq:CBverge}
\end{align}
If \(C_{\rm B,\pm}\) turns imaginary, the whole \(C_{\rm B}\) range is allowed \( C_{\rm B} \in \left[-1, 1\right]\) in accordance to Eq. \ref{eq:EBvergeDetgeq0CB}. 

The integration of the elliptic phase space volume Eq. \ref{eq:3ddist} is preformed over the intersection of the needed domain (e.g.  \ref{eq:integrationBoundsNaive}), with the domains described in this section.

\section{Analytic approximation of binary eccentricity distribution}
\label{app:eBapproximation}

\begin{figure*}
\centering
\includegraphics[width=80mm]{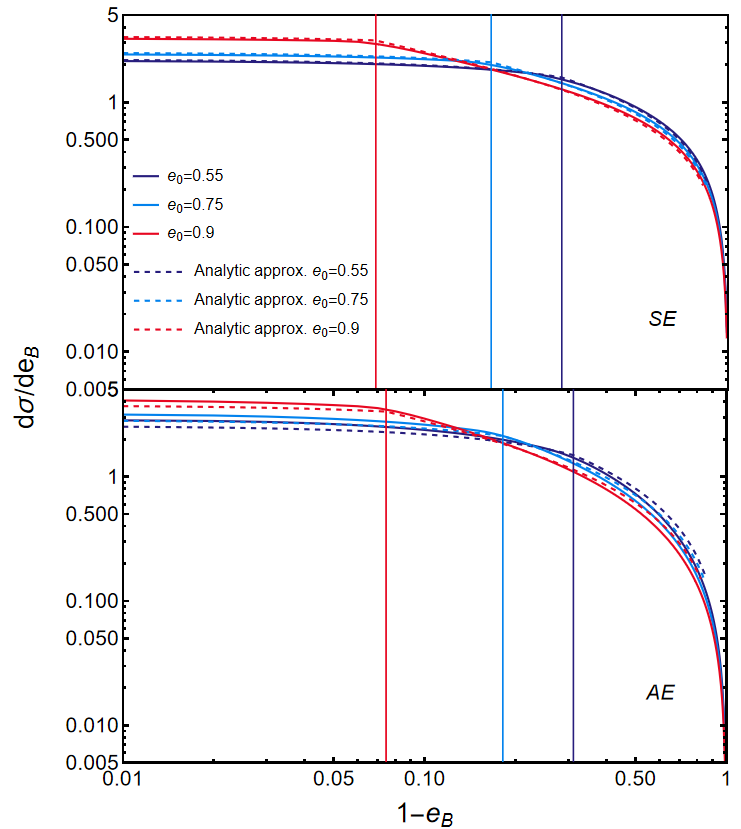}
\caption{The elliptic eccentricity distribution is plotted against \(1 - e_{\rm B}\). The distribution assumes equal masses. 
Full lines are the exact (numerically marginalized) distribution and the dashed lines are analytic approximations to the distributions given in Eqs. \ref{eq:eBapproxSESpike}-\ref{eq:eBapproxAESpike}. The vertical lines correspond to the location of the 2nd order discontinuity given in Eq. \ref{eq:Spike}. Different \(e_{\rm 0}\) values are given in different colors; 0.55 in dark blue, 0.75 in light blue and 0.9 in red. The top panel examines the SE criteria for the chaotic region and the bottom panel examines the AE criteria.  Generally, the fully analytic approximations of Appendix \ref{app:eBapproximation} provide a good fit to more exact calculations.}
\label{fig:eBapproximation}
\end{figure*}

Inspecting the elliptic eccentricity distribution \(d\sigma/de_{\rm B}\) presented in Fig. \ref{fig:eBdistDividedtoAESE}, we can spot a 2nd order discontinuity for intermediate \(e_{\rm 0}\) values. We connect this 2nd order discontinuity to the 2d divergence given in Eq. \ref{eq:Spike} and denote it by \(e_{\varcurlywedge}\). We approximate the marginal distribution of intermediate state eccentricities by noting that, before the divergence at \(e_{\rm B} \leq e_{\varcurlywedge}\), the distribution is strongly super-thermal, and that after the divergence \(e_{\varcurlywedge} \leq e_{\rm B}\), the distribution becomes thermal or approximately so. Demanding normalization and continuity at \(e_{\rm B} = e_{\varcurlywedge}\), we obtain the following analytic approximations:
\begin{align}
    \left(SE\right):\,\,\,\, \frac{d\sigma}{de_{B}}=
    \begin{cases}
        \left(\frac{1}{1-\frac{1}{2}\sqrt{1-e_{\varcurlywedge}^{2}}}\right)\frac{e_{B}}{\sqrt{1-e_{B}^{2}}} & e_{B}<e_{\varcurlywedge}\\
        \left(\frac{1}{\sqrt{1-e_{\varcurlywedge}^{2}}-\frac{1}{2}\left(1-e_{\varcurlywedge}^{2}\right)}\right)e_{B} & e_{B}>e_{\varcurlywedge}
    \end{cases}
    \label{eq:eBapproxSESpike}
\end{align}
for SE chaotic region criteria, and
\begin{align}
    \left(AE\right):\,\,\,\, \frac{d\sigma}{de_{B}}=
    \begin{cases}
        C_{1}\frac{e_{B}\left(1+e_{B}\right)}{\sqrt{1-e_{B}^{2}}} & e_{B}<e_{\varcurlywedge}\\
        C_{1}\frac{e_{B}\left(1+e_{B}\right)}{\sqrt{1-e_{\varcurlywedge}^{2}}} & e_{B}>e_{\varcurlywedge}
    \end{cases}
    \label{eq:eBapproxAESpike}
\end{align}
for AE chaotic region criteria, where 
\begin{align}
    C_{1}=\left[\frac{1}{2}\sin^{-1}\left(e_{\varcurlywedge}\right)-\frac{1}{2}\left(e_{\varcurlywedge}+2\right)\sqrt{1-e_{\varcurlywedge}^{2}}+1+\frac{\left(5-2e_{\varcurlywedge}^{3}-3e_{\varcurlywedge}^{2}\right)}{6\sqrt{1-e_{\varcurlywedge}^{2}}}\right]^{-1}
\end{align}
is the normalization constant.
In Fig. \ref{fig:eBapproximation} we plot the elliptic eccentricity distribution for mid-range \(e_{\rm 0}\) triples along with the analytic approximations. We use the expected value of the binary energy to evaluate \(e_{\varcurlywedge}\) with Eq. \ref{eq:Spike} and find good agreement in the SE case. It is worth noting that a better agreement was reached in the AE case when taking \(E_{\rm B}\) in Eq. \ref{eq:Spike} to be \(E_{\rm cut}\) as in Eq. \ref{eq:EBcut_def} instead of the expected value, but it was slightly worse for the SE case.

\section{Scramble count statistics and parameter dependence}
\label{app:beta}

In this Appendix, we examine how the triple lifetime distribution (examined here in dimensionless form as a scramble count distribution) depends on the two fudge factors of our intermediate states formalism: the free parameters $\alpha$ (which, as in \citet{StoneLeigh19}, describes the physical size of the strong/chaotic interaction region) and $\beta$, a parameter that quantifies the transition between a scramble and a tertiary excursion during an intermediate state. We focus on the simple escape criteria for the chaotic region as given in Eq. \ref{eq:simpleesc}.

\begin{figure}
     \centering
     \begin{subfigure}{0.45\textwidth}
         \centering
         \includegraphics[width=\linewidth]{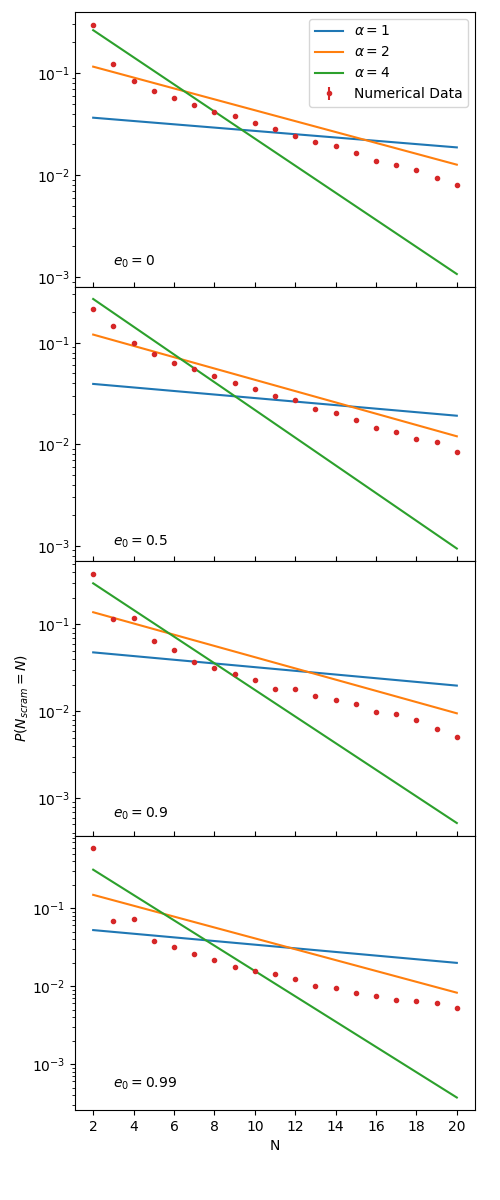}
         \caption{Variations in $\alpha$, the physical size of the chaotic region. \\  Analytic predictions (lines) assume the SE criterion.}
         \label{fig:alphasvsNumerical}
     \end{subfigure}
     \begin{subfigure}{0.45\textwidth}
         \centering
         \includegraphics[width=\linewidth]{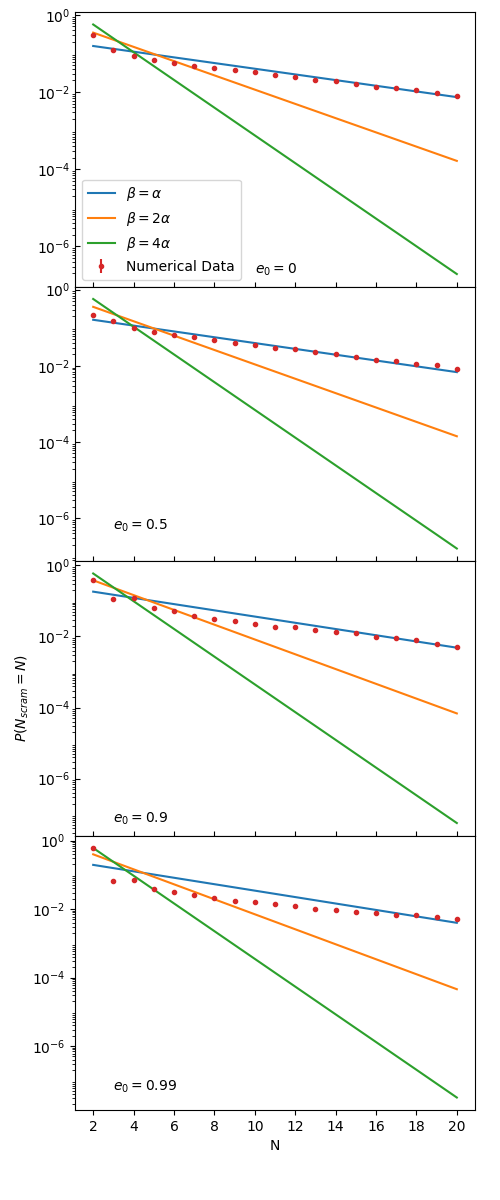}
         \caption{Variations in $\beta$, the boundary between a chaotic scramble and a temporarily hierarchical excursion.}
         \label{fig:bettasvsNumerical}
     \end{subfigure}
     \caption{Distributions of scramble count $N$ from our analytic models and numerical simulations.  Different rows show different numerical ensembles of binary-single scattering events labeled by angular momentum deficit parameter $e_0$.  Numerical data taken from \citealt{StoneLeigh19} are shown as red dots, and the sum of these numerical probabilities are normalized to 1, starting at \(N_{\rm scram} = 2\).  The analytic probability of an SE system to exhibit a certain number of scrambles $N$ is plotted against $N$ in colored lines.  {\it Column (a)}: calculations with \(\alpha = 1\) are shown in blue, \(\alpha = 2\) in orange, and \(\alpha = 4\) in green. All panels in this column assume \(\beta = \alpha\). We see that across all angular momenta, the best fit value for \(\alpha\) lies between 2 and 4, with $\alpha\approx 2.5$ roughly providing the best match.  {\it Column (b)}: the same as the previous column, but now $\alpha=2.5$ is fixed and the analytic predictions vary \(\beta\). Calculations with \(\beta = \alpha\) are shown in blue, \(\beta = 2 \alpha\) in orange, and \(\beta = 4 \alpha\) in green. The best agreement with the data is achieved when \(\alpha=\beta\).}
     \label{fig:1}
\end{figure}

In Figs. \ref{fig:alphasvsNumerical} and \ref{fig:bettasvsNumerical}, we show both analytic predictions for the scramble count distribution $P(N_{\rm scram})$ as well as the outcome of numerical scattering experiments.  The numerical data, which is taken from \citet{StoneLeigh19}, is broken down into four separate ensembles.  Each ensemble has a different $L_0$ value (though all have the same $E_0$ and mass triplet).  We see that the scramble count distribution is best reproduced with $\alpha \approx 2.5$ and $\beta \approx \alpha$, indicating that the ergodic formalism for intermediate and final state formation can be well described with just one free parameter, rather than two.

\begin{figure*}
\centering
\includegraphics[width=100mm]{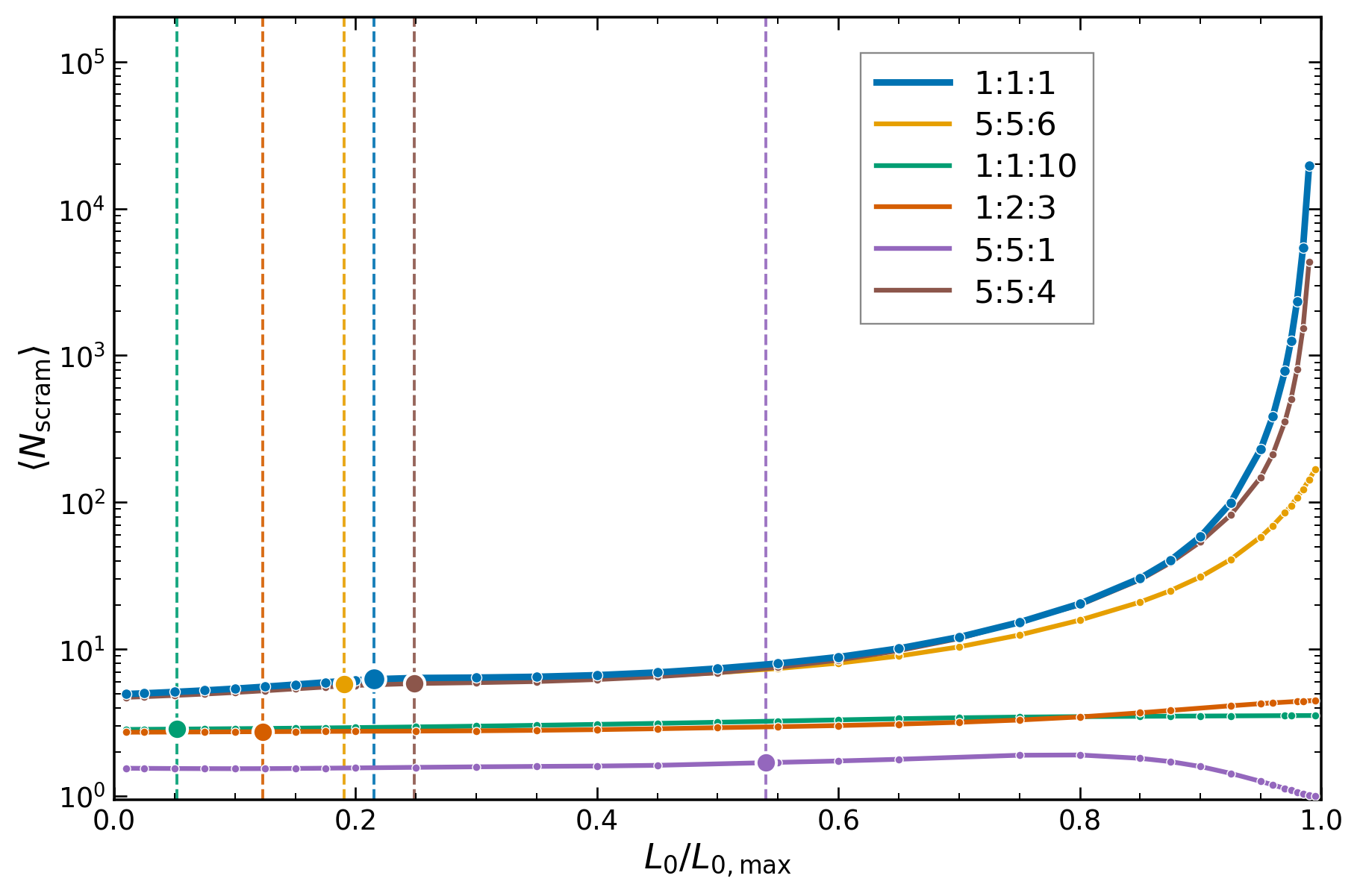}
\caption{Expected number of scrambles,
$\left\langle N_{\rm scram}\right\rangle=p_{\rm dis}^{-1}$,
as a function of the normalized initial angular momentum
$L_0/L_{0,\max}$ for the mass ratios indicated in the legend ($m_1/M:m_2/M:m_3/M$).
The disintegration probability $p_{\rm dis}$ is calculated from the
phase space (Eq. \ref{eq:Pdis}) model with $\alpha=\beta=2.5$.
The total mass is fixed at $M=60\,M_\odot$, and the initial binary
comprises $m_1$ and $m_2$.
For each mass triplet, the total energy is set to
$E_0=-Gm_1m_2/\left(2a_{\rm ref}\right)$ with
$a_{\rm ref}=1\,\mathrm{AU}$; the angular momentum is normalized by
the corresponding cutoff $L_{0,\max}$ for this initial binary.
The colored vertical dashed lines mark circular binary reference
states with $E_B=E_0$, $e_{B,0}=0$, and zero outer orbital angular
momentum, $L_s=0$.
The matching circles show the corresponding values of
$\left\langle N_{\rm scram}\right\rangle$. We see that as the total angular momentum rises for the equal mass case, the number of scrambles can increase across a few orders of magnitude.}
\label{fig:mean_scrambles_L0}
\end{figure*}

Having adopted $\alpha=\beta=2.5$, we examine the resulting mean
number of scrambles before escape. At fixed total energy, angular
momentum, and component masses, the geometric distribution in
Eq.\ref{eq:PofNscram} gives
$\left\langle N_{\rm scram}\right\rangle=p_{\rm dis}^{-1}$.
Figure~\ref{fig:mean_scrambles_L0} illustrates the dependence of
this mean on angular momentum for several representative mass
triplets, with each curve normalized by its corresponding
$L_{0,\max}$ given in Appendix \ref{app:L0maxder}. 
The sharp increase in $\left\langle N_{\rm scram}\right\rangle$ near $L_{0,\max}$ in the equal mass case results from the suppression of the escaping phase space volume. As $L_0$ approaches this boundary, escaping configurations become increasingly restricted, requiring nearly circular binaries, aligned orbital angular momenta, and nearly parabolic escaping trajectories grazing the boundary of the chaotic region. At $L_{0,\max}$, these conditions can only be satisfied by a limiting configuration of vanishing phase space volume, while bound excursions remain accessible, causing $p_{\rm dis}\to 0$. For unequal masses, however, the three possible binary--escaper combinations have different angular momentum cutoffs. Consequently, other escape channels may remain accessible at the $L_{0,\max}$ defined for the initial binary, preventing $p_{\rm dis}$ from vanishing at that boundary.

\section{Derivation of \(L_{\rm 0, max}\)}
\label{app:L0maxder}

Consider a binary \(m_{\rm a}, m_{\rm b}\) with angular momentum \(\vec{L}_{\rm B}\) and an incoming single \(m_{\rm s}\) with angular momentum \(\vec{L}_{\rm s}\). The total angular momentum of the encounter is then
\begin{align}
    L_{\rm 0}^2 = L_{\rm B}^2 + L_{\rm s}^2 +2L_{\rm B} L_{\rm s} C_{\rm B} \leq \left(L_{\rm B} + L_{\rm s}\right)^2,
    \label{eq:L0ofLBLsCB}
\end{align}
where \(C_{\rm B} \in \left[-1, 1\right]\) is the cosine of the inclination, which maximizes \(L_{\rm 0}\) at \(C_{\rm B} = 1\).

If the single body has a large enough impact parameter (or equivalently a large enough angular momentum) when approaching the binary, the 3 body system will remain hierarchical and there will be no formation of a chaotic triple (\citealt{valtonen_three-body_2006} chapter 7.5). We can construct an upper bound on \(L_{\rm 0}\) by expressing upper bounds on \(L_{\rm B}, L_{\rm s}\). The binary angular momentum \(L_{\rm B}\) is maximized for a circular orbit:
\begin{align}
    L_{\rm B} = \mathcal{M}\sqrt{G m_{\rm B} a_{\rm B}\left(1 - e_{\rm B}^2\right)} \leq \mathcal{M}\sqrt{G m_{\rm B} a_{\rm B}}.
    \label{eq:LBUpperboundL0max}
\end{align}
For the tertiary angular momentum \(L_{\rm s}\), we argue that for a chaotic triple to emerge, the tertiary's pericenter needs to be inside the chaotic region, in accordance with Eq. \ref{eq:qslessR}:
\begin{align}
    L_{\rm s} = m\sqrt{G M a_{\rm s} \left(e_{\rm s}^2 -1\right)}\leq m\sqrt{G M R \left(2 + \frac{R}{a_{\rm s}}\right)}
    \label{eq:LsUpperboundL0max}
\end{align}
Plugging Eqs. \ref{eq:LsUpperboundL0max}, \ref{eq:LBUpperboundL0max} into Eq. \ref{eq:L0ofLBLsCB} and converting the semi-major axes to energies we get an upper bound on \(L_{\rm 0}\) that depends on the binary energy:
\begin{align}
    L_{0}\leq\mathcal{M}\sqrt{-\frac{G^{2}m_{B}m_{a}m_{b}}{2E_{B}}}+m\sqrt{2GMR\left(1+\frac{E_{s}R}{Gm_{s}m_{B}}\right)}. 
\end{align}
Using the simple escape criterion from Eq. \ref{eq:simpleesc} with energy conservation and rearranging for the explicit \(E_{\rm B}\) dependence, we find
\begin{align}
    L_{\rm 0}\leq\sqrt{\frac{G^{2}m_{\rm B}\mathcal{M}}{\left(-E_{\rm B}\right)}}\left(\mathcal{M}\sqrt{\frac{m_{\rm B}}{2}}+m\sqrt{\alpha M\left(1+\alpha\frac{m_{\rm B}\mathcal{M}\left(E_{\rm 0}-E_{\rm B}\right)}{2Mm\left(-E_{\rm B}\right)}\right)}\right).
    \label{eq:L0maxEBalpha}
\end{align}
 The energy of the incoming single is positive and therefore \(E_{\rm 0} > E_{\rm B}\). We now notice that the maximum over \(E_{\rm B}\) for the right hand side of Eq. \ref{eq:L0maxEBalpha} is obtained when \(E_{\rm B} = E_{\rm 0}\).
The common upper bound for any \(E_{\rm B}, C_{\rm B}, e_{\rm B}\) is then:
\begin{align}
    L_{\rm 0, max} = \sqrt{\frac{G^{2}m_{\rm a}m_{\rm b}}{\left(-E_{\rm 0}\right)}}\left(\mathcal{M}\sqrt{\frac{m_{\rm B}}{2}}+m\sqrt{\alpha M}\right) = G \sqrt{- \frac{1}{E_{\rm 0}}} \mu^{\frac{5}{2}},
    \label{eq:L0maxE0}
\end{align}
where we denote the mass dependency
\begin{align}
    \mu^{\frac{5}{2}} = \sqrt{m_{\rm a} m_{\rm b}} \left(\mathcal{M}\sqrt{\frac{m_{\rm B}}{2}}+m\sqrt{\alpha M}\right)
    \label{eq:Mscr}
\end{align}
for short.

Using the value of \(\alpha = 2.5\) suggested by our numerical scattering experiments and specializing to the equal mass case \(m_{\rm a} = m_{\rm b} = m_{\rm s} = \tilde{m}\), we get
\begin{align}
    L_{\rm 0, max} \approx 2.33 G \sqrt{-\frac{\tilde{m}^5}{E_{\rm 0}}},
\end{align}
differing only slightly from the result in \citealt{valtonen_three-body_2006} Eq. 7.27, in which a somewhat different argument finds \(L_{\rm 0, max} = 2.5 G \sqrt{-\tilde{m}^5/E_{\rm 0}}\).

\section{Derivation of \(M_{\rm ej}\)}
\label{app:Mejder}
Consider a scenario where the semi-major axis for a FS merger \(a_{\rm GW}\) becomes larger than the one needed for a binary to be ejected from the cluster \(a_{\rm ej}\), i.e. $a_{\rm ej} \le a_{\rm GW}$.  
While $a_{\rm GW}$ generally depends on binary eccentricity, we choose here the $e_{\rm B}=0$ limit, $a_{\rm GW} = (4\eta T_{\rm bs})^{1/4}$ (where \(\eta\) is as defined in Eq. \ref{eq:eta}), as this represents the slowest GW inspiral times and is thus the most optimistic case for binary ejections. Re-writing Eq. \ref{eq:EnergykickForEj} in terms of the binary semi-major axis, we thus obtain the ejection semi-major axis
\begin{align}
    a_{\rm ej} = - \frac{G m_{\rm a} m_{\rm b}}{2\left(E_{\rm 0} - \frac{m_{\rm B}}{2}\left(1 + \frac{m_{\rm B}}{m_{\rm s}}\right)v_{\rm esc}^2\right)}.
    \label{eq:aEj}
\end{align}
Assuming the expected value of binary energy \(E_{\rm B}\) is an order unity multiple of the total triple energy \(E_{\rm 0}\), i.e.,
\begin{align}
    \left<E_{\rm B}\right> = k E_{\rm 0} 
    \label{eq:ExpEBiskE0}
\end{align}
then for some constant \(k \geq 1\),
we can eliminate the explicit $E_0$ dependence in Eq. \ref{eq:aEj}: 
\begin{align}
    \left<a_{\rm ej}\right> = \frac{G m_{\rm a} m_{\rm b}}{\left(\frac{k}{k-1}\right)m_{\rm B}\left(1 + \frac{m_{\rm B}}{m_{\rm s}}\right)v_{\rm esc}^2}.
    \label{eq:aEjwithk}
\end{align}
Now we combine Eq. \ref{eq:aEjwithk} with the circular-orbit limit of $a_{\rm GW}$, to find a maximum $M_{\rm tot, \star}$ for which ejections remain possible:
\begin{align}
    &M_{\rm ej} \equiv \left(\frac{5^{\frac{1}{5}}\pi^{\frac{9}{10}}\left(1-\frac{1}{k}\right)}{2^{\frac{41}{10}}3^{\frac{1}{10}}\left(\frac{\pi^{2}}{8}-1\right)^{\frac{1}{10}}}\frac{c}{G^{\frac{1}{2}}}\frac{m_{a}^{\frac{4}{5}}m_{\rm b}^{\frac{4}{5}}\left\langle m_{s}\right\rangle ^{\frac{4}{5}}}{m_{\rm B}^{\frac{6}{5}}\left(m_{\rm B}+\left\langle m_{\rm s}\right\rangle \right)^{\frac{4}{5}}}r_{\rm h}^{\frac{1}{2}}\frac{\left(1+c_{\circ}\right)^{\frac{1}{5}}\left(c_{\circ}-1\right)}{\ln\left(c_{\circ}\right)c_{\circ}^{\frac{4}{5}}}\right)^{\frac{10}{9}} .
    \label{eq:Mej}
\end{align}
Here \(c\) is the speed of light, \(c_{\rm \circ}\) is the cluster's concentration parameter, \(r_{\rm h}\) is the cluster's half mass radius, and \(\left\langle m_{\rm s}\right\rangle\) is the average mass on an incoming single object. Using the half radius-mass relationship from Eq. \ref{eq:rhOfMtot}, we obtain
\begin{align}
    &M_{\rm ej} \simeq M_{\odot}^{\frac{5}{5+2s}} \times \left(10^{4.67}\times\left(1-\frac{1}{k}\right) \frac{m_{\rm a}^{\frac{4}{5}}m_{\rm b}^{\frac{4}{5}}\left\langle m_{\rm s}\right\rangle ^{\frac{4}{5}}}{m_{\rm B}^{\frac{6}{5}}\left(m_{\rm B}+\left\langle m_{\rm s}\right\rangle \right)^{\frac{4}{5}}}\frac{\left(1+c_{\circ}\right)^{\frac{1}{5}}\left(c_{\circ}-1\right)}{\ln\left(c_{\circ}\right)c_{\circ}^{\frac{4}{5}}}\right)^{\frac{5s}{5+2s}}.
    \label{eq:Mejrhcallib1}
\end{align}
Plugging in \(k\simeq \frac{3}{2}\), \(c_{\circ} = 3.5\) and \(\left\langle m_{\rm s}\right\rangle = m_{\rm a} = m_{\rm b} = 20 M_{\odot}\) we get \(M_{\rm ej} \simeq 10^{6.20} M_{\odot}\). 
Correcting the derivation for 2 component clusters, by starting from Eq. \ref{eq:Mej} and using the \(r_{\rm h,BH} \left(r_{\rm h, \star}\right)\), \(M_{\rm tot, BH} \left(M_{\rm tot, \star}\right)\) relations along with the half-mass radius - total mass relationship for the stellar parent cluster from Eq. \ref{eq:rhOfMtot}, we obtain Eq. \ref{eq:Mejrhcallib}.

\section{Merger outcome plots}
\label{app:moreMCresults}
\begin{figure*}
\includegraphics[width=160mm]{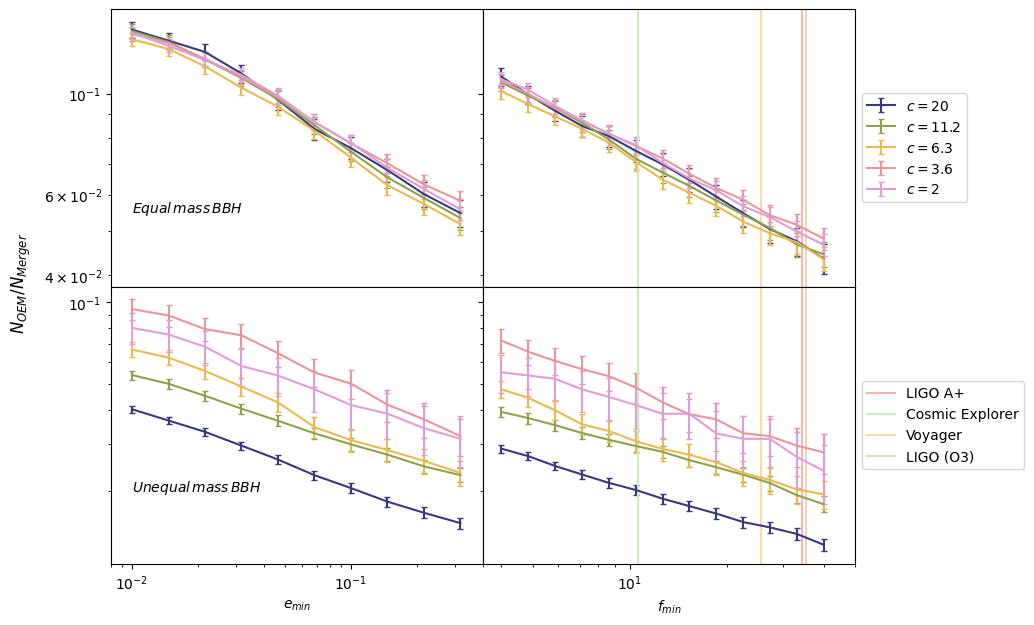}
\caption{Ratios of sampled eccentric merger dissolutions to the total number of sampled mergers as a function of detection sensitivity. In the left panels, minimal eccentricity \(e_{\rm f}\) for detection at frequency \(f = 10\) Hz is varied, while in the right panels the frequency for detection is varied with eccentricity \(e_{\rm f} = 0.1\). The top panels present results of the unequal mass BBH case, and the bottom present the equal mass BBH case. Different curves depict different cluster concentration parameters. Vertical lines corresponding to different detectors' sensitivities were added to the right panels. We can see that the fraction of eccentric mergers to total mergers is in the 0.04 - 0.14 range for the equal mass BBH case and 0.01 - 0.07 for the unequal mass BBH case.}
\label{fig:NEM2MergerCsGrid}
\end{figure*}
\twocolumn
\begin{figure}
\includegraphics[width=80mm]{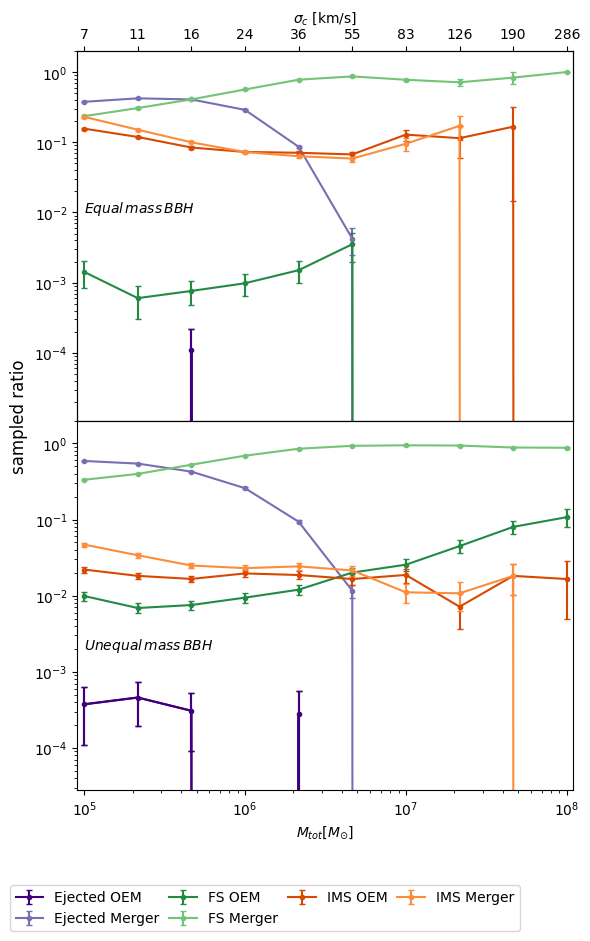}
\caption{Ratios of sampled BBH merger dissolutions plotted against varying star cluster total mass measured in solar masses, and half-mass radius according to Eq. \ref{eq:rhOfMtot}. The top panel corresponds to the equal mass BBH case while the bottom panel corresponds to the unequal mass BBH case. The ratio of each point was taken such that the sum of all vertical points for each \(M_{\rm tot}\) value is 1. In dark orange IMS OEM are shown, in orange IMS non-eccentric mergers, in dark green FS OEM, in green non-eccentric FS mergers, in purple ejected non-eccentric mergers and in dark purple ejected OEM.}
\label{fig:ratios_Mcs_merger_only}
\end{figure}
\begin{figure}
\includegraphics[width=80mm]{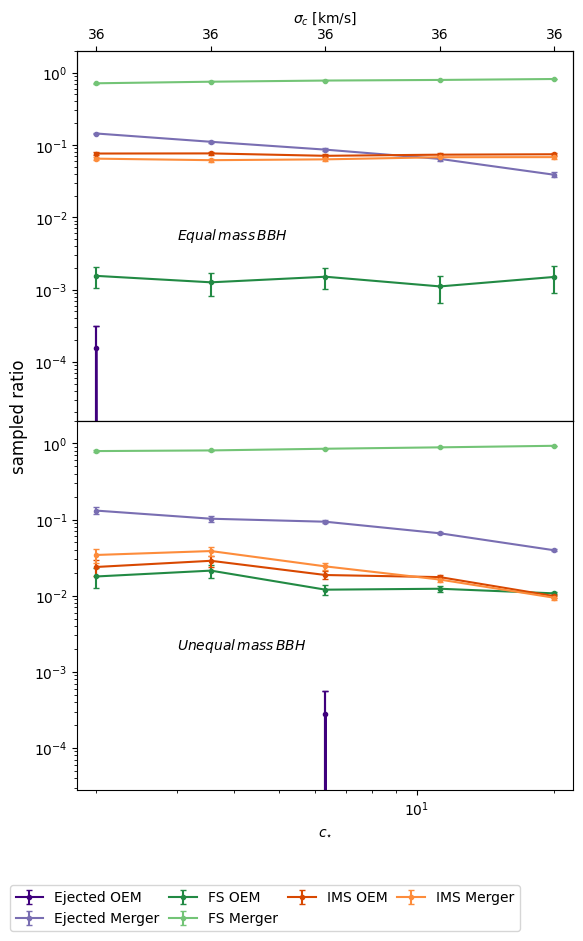}
\caption{Ratios of sampled BBH merger dissolutions plotted against varying star cluster concentration parameter \(c_{\rm \star}\). The top panel corresponds to the equal mass BBH case while the bottom panel corresponds to the unequal mass BBH case. The ratio of each point was taken such that the sum of all vertical points for each \(c_{\star}\) value is 1. In dark orange IMS OEM are shown, in orange IMS non-eccentric mergers, in dark green FS OEM, in green non-eccentric FS mergers, in purple ejected non-eccentric mergers and in dark purple ejected OEM.}
\label{fig:ratios_cs_merger_only}
\end{figure}
\onecolumn

\label{lastpage}
\end{document}